\newlength{\intwidth}

\documentclass{jfm}

\usepackage{amssymb}

\usepackage{array}
\usepackage{amsmath}
\usepackage{latexsym}
\usepackage{bezier}
\usepackage{psfrag,graphicx}
\usepackage{bm}
\usepackage{float}
\usepackage{multirow}
\usepackage{mathrsfs}
\usepackage{color}
\usepackage{enumerate}
\usepackage{natbib}
\usepackage{nicefrac}

\newcommand{\RealN}{\mathbb{R}}

\newtheorem{thm}{Theorem}

\begin{document}

\title[Hurdle theorem applied to alternating jets]{\textcolor{black}{A sufficient condition for inviscid shear instability: Hurdle theorem and its application to alternating jets}
}



\author{
  K. Deguchi\aff{1}\corresp{\email{kengo.deguchi@monash.edu}},
  M. Hirota\aff{2}
  \and T. Dowling\aff{3}
}
 
 \affiliation{
   \aff{1}School of Mathematics, Monash University, VIC 3800, Australia
   \aff{2}Institute of Fluid Science, Tohoku University, Sendai, 980-8577, Japan
   \aff{3}Department of Physics \& Astronomy, University of Louisville, KY 40299 USA
}

\maketitle

\begin{abstract}

We propose a simple method to identify unstable parameter regions in general inviscid unidirectional shear flow stability problems. The theory is applicable to a wide range of basic flows, including those that are non-monotonic. We illustrate the method using a model of Jupiter's alternating jet streams based on the quasi-geostrophic equation. The main result is that the flow is unstable if there is an interval in the flow domain for which the reciprocal Rossby Mach number \textcolor{black}{(a quantity defined in terms of the zonal flow and potential vorticity distribution),} surpasses a certain threshold or `hurdle'. The hurdle height approaches unity when we can take the hurdle width to greatly \textcolor{black}{exceeds} the atmosphere's intrinsic deformation length, as holds on gas giants. 
\textcolor{black}{In this case, the Kelvin-Arnol'd sufficient condition of stability accurately detects instability. 
}
\textcolor{black}{These results improve the theoretical framework for explaining the stable maintenance of Jupiter and Saturn's jets over decadal timescales.}

\end{abstract}

\section{Introduction}
The stability of inviscid, parallel shear flows has applications in geophysical fluid dynamics, astrophysics, plasma physics, and engineering thermofluid sciences. For the purely hydrodynamic cases, sufficient conditions for stability can be traced back to the \cite{Rayleigh1880} inflection-point theorem, the refinement of which by \cite{Fjortoft1950} \textcolor{black}{was later revealed to be} one of the two conditions derived using Arnol'd's method \citep{Arnold1966}. The existence of two distinct sufficient conditions was anticipated by Kelvin on energetic grounds \citep{Thomson1880}, the same year as Rayleigh's inflection point theorem. Today they are called the Kelvin-Arnol'd $1^{\rm st}$ and $2^{\rm nd}$ shear-stability theorems (KA-I and KA-II). KA-I corresponds to the Rayleigh-Fj{\o}rtoft condition, while KA-II is relevant in planetary physics.

A breach of sufficient conditions for stability does not necessarily imply instability. 
However, in applied fields, these conditions are sometimes treated as if they were sharp stability criteria, \textcolor{black}{meaning that they accurately detect the stability boundary,} contributing to potential confusion.
The reason for this may be that the known necessary conditions for stability, i.e., conditions guaranteeing the existence of instability, require rather complex assumptions about the basic flow, $U(y)$. \cite{Tollmien1935}
was the first to find a class of basic flows where the KA-I condition becomes a sufficient and necessary condition for stability. Many assumptions are required of $U$ in his class, e.g. $U$ is symmetric and disappears on the wall, etc. \cite{Howard1964} also showed that KA-I becomes a sharp stability condition when the flow is considered in an unbounded domain and the basic flow is assumed to be in the class $\mathcal{H}$ he defined. Nevertheless the above results are only special cases and in general KA-I is not sharp. In a bounded domain, it is in fact possible to construct counterexamples that demonstrate that KA-I is not a necessary condition of stability \citep[see][]{Tollmien1935, Drazin_Howard1966, Drazin_Reid1981}.

The more general the basic flow considered, the more complex the argument required to derive necessary conditions for stability. Using the Nyquist method, \cite{Rosenbluth_Simon1964} derived a sharp stability condition using an integral quantity that can be calculated from the basic flow. The assumption imposed by them on $U(y)$ is that it is monotonic and has only one inflection point in the domain. The latter assumption was removed by \cite{Balmforth_Morrison1999}, who also used the Nyquist method. \cite{Hirota_etal2014}
used the variational method under almost the same assumptions and demonstrated that the definiteness of the quadratic form introduced by \cite{Barston1991} is a necessary and sufficient condition for stability. In order to derive those conditions, it is critical to assume the monotonicity of $U(y)$. The special nature of this type of basic flow is that all neutral modes must possess only one critical layer---where the phase speed of the mode matches $U(y)$---and moreover, the position of this layer is fixed at one of the inflection points.

Demonstrating the existence of a neutral mode first and then perturbing the wavenumber to find unstable modes has been a common approach since \cite{Tollmien1935} and \cite{Howard1964}. The assumptions employed by \cite{Howard1964} were aimed at enabling the use of Sturm-Liouville theory for neutral modes \citep[see also][]{Morse_Feshbach1953}, allowing the assertion of the existence of neutral modes for non-monotonic $U$. For the bounded flows, \cite{Tung1981}
investigated how much the assumptions about $U$ can be relaxed to demonstrate the presence of unstable modes around neutral modes. While his analysis had some flaws, later, \cite{Lin2003}
independently completed a mathematically rigorous theory.

One notable example where the non-monotonicity of the basic velocity field becomes crucial is the alternating jet streams of Jupiter and Saturn. Their evolution is governed by the behaviour of Rossby waves, which are analogous to drift waves in plasma physics and arise when there is a gradient in the potential vorticity (PV), a conserved fluid-dynamical quantity formed by the combination of conservation of mass, momentum, and thermal energy \citep{Vallis2017}. Following standard practice, this article works within the quasi-geostrophic framework, which admits Rossby waves but filters out sound waves and \textcolor{black}{inertia-gravity waves.} The non-rotating, non-stratified KA-I result by Rayleigh and Fj{\o}rtoft was followed by extensions to rotating, non-stratified flow \citep{Kuo1949} and to rotating, stratified flow \citep{Charney1962}. The latter result, widely known as the Charney-Stern stability criterion, established that a sufficient condition for stability is the absence of a point where the PV gradient changes sign, \textcolor{black}{which we refer to as \textit{PV extremum.} Note that in geophysics, a PV extremum is sometimes referred to as a \textit{critical latitude} because the observed phase speed of the mode often matches $U(y)$ there. In this paper, however, a clear distinction is made between the two.}



\textcolor{black}{In terms of the basic state streamfunction, $\Psi$, \textcolor{black}{defined by $U=-d \Psi/dy$,} and the associated PV, $Q$, the KA-I and KA-II conditions can be expressed as \textcolor{black}{$-dQ/d\Psi \leq 0$ and $0\leq -dQ/d\Psi \leq l^{-2}$,} respectively, when perturbations have 
\textcolor{black}{the largest length scale $l$}
 \textcolor{black}{\citep{McIntyreShepherd1987}}.}
Nondimensionalising \textcolor{black}{those} 
 conditions for the planetary atmosphere problem uncovers the role played by the Rossby Mach number \textcolor{black}{defined by 
 \begin{eqnarray}
M=\frac{\kappa_0^2(U-\alpha)}{Q'}, \label{RosM0}
\end{eqnarray}
where $\kappa_0$ is a constant determined by the length scale of the eddies and $\alpha$ is a suitable Galilean shift.}
\textcolor{black}{\cite{Dowling2014,Dowling2020} introduced $M$ based on physical intuition applied to the net propagation of the fastest Rossby waves relative to the flow, which are the longest waves. In brief, KA-I and KA-II can be interpreted as establishing that unstable flows must become ``subsonic'' (i.e. $0< M<1$) somewhere, but with respect to Rossby waves rather than sound waves. }
Getting KA-I and KA-II to concatenate together via $M^{-1}\leq 1$ is a compact way to look at the stability condition.
The above simple stability condition can only be applied for basic flows belonging to a certain class where the sign of $M^{-1}(y)$ does not change for all $y$. This class, in fact, closely resembles the one defined by \cite{Howard1964} and \cite{Lin2003}, and we base our analysis on this ground.

\begin{figure}
\begin{center}
  \includegraphics[width=0.7\textwidth]{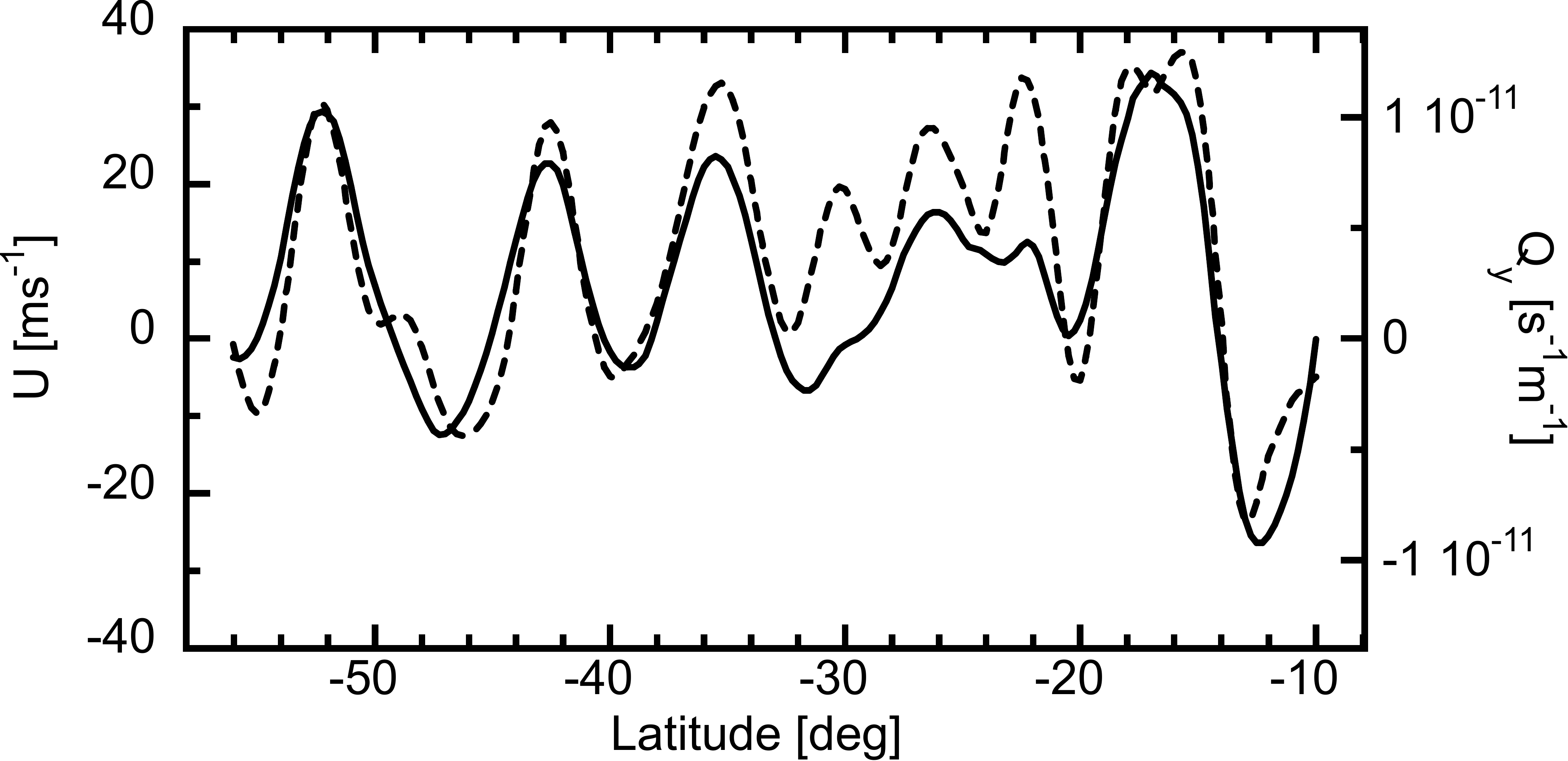}
\end{center}
\caption{
Example Jupiter profiles with respect to latitude of zonal wind, $U$ (solid line), and potential vorticity gradient, $Q' = dQ/dy = Q_y$ (dashed line), from the southern hemisphere, at 11 hPa pressure. A strong correlation between $U$ and $Q_y$ is evident. Here, $dy = R d\phi$, where $y$ is latitude in meters, $\phi$ is latitude in radians (planetographic), and $R(\phi)$ is the local radius of curvature \citep[see equation (5) in][]{Dowling1989}.  These are a sample of the profiles produced by \cite{Read2006} using the Cassini Composite Infrared Spectrometer (CIRS). \textcolor{black}{Similar correlations to this figure are found at other pressure levels.}
}
\label{fig:Read2006_profiles}
\end{figure}

\cite{Dowling1993} discovered that Jupiter's atmosphere has $U$ and $Q'$ profiles strongly correlated. This analysis, based on \textit{Voyager} observations of the cloud-top vorticity field \citep{Dowling1989}, was expanded for Jupiter and Saturn by \cite{Read2006} and \cite{Read2009,Read2009b}, as illustrated in figure~\ref{fig:Read2006_profiles}. The correlation implies that the $M^{-1}(y)$ profile is almost constant. If the profile is neutrally stable and the KA-II condition is sharp, this constant must be unity. However, as already noted, the stability condition based on the reciprocal Rossby Mach number is only a sufficient condition for stability, and the significance of it regarding necessary conditions for stability remains not well understood. 
\textcolor{black}{Moreover, the observed $M^{-1}(y)$ profiles are not precisely constant,} 
as can be seen from the fact that the correlation in figure~\ref{fig:Read2006_profiles} is not perfect.
\textcolor{black}{The neutrally stable hypothesis for Jupiter’s jets has been studied in the KA-II context for more than 30 years, as summarised in the most recent review article by \cite{Read2024}, but a conclusive resolution has yet to be reached.}

Particularly intriguing from a planetary atmospheric physics perspective is thus when KA-II gives sharp stability boundaries. There are interesting numerical results in this regard: \cite{Stamp1993} set up a sinusoidal model basic flow such that $M$ becomes a constant, and numerically found that the instability disappears at the KA-II stability boundary, $M=1$. Motivated by these empirical results, our mathematical goals are twofold: 
(i) to derive \textcolor{black}{simple} conditions that guarantee the presence of instability, and 
(ii) to determine under what conditions KA-II achieves sharp stability boundaries. We also attempt to emphasise the simplicity of the theoretical results, because the pursuit of sharpness of the conditions often makes them difficult to use. For example, from a practical standpoint, the condition proposed by \cite{Balmforth_Morrison1999} is not \textcolor{black}{easily applicable} due to the requirement of solving a Fredholm integral equation. Also, to demonstrate the existence of instability using the quadratic form in \cite{Hirota_etal2014} or the Rayleigh quotient in \cite{Lin2003}, it is necessary to find a convenient test function, but no useful recipe is known for non-monotonic basic flows in a finite domain. 

\textcolor{black}{
The rest of the article is organised as follows. In Section \ref{sec:formulation}, the quasi-geostrophic equation is linearised around the basic flow to yield an eigenvalue problem. Our sufficient condition for instability, in the form of a hurdle of the $M^{-1}(y)$ profile, is stated in Section 3, together with the KA stability theorem. Basic states are classified in the same section to clarify when the stability conditions can be used. One of the classes is identified as having Jupiter-style critical latitudes; it yields an almost sharp stability condition that aligns with observations of the jets on Jupiter and Saturn. In Section 4, the stability criteria are illustrated with model basic flow profiles. Section 5 studies the implications of the theoretical results obtained in the previous two sections to planetary physics problems. Section 6 contains mathematical proofs of the article's theorems. Our strategy is to extend the Rayleigh quotient method, developed by \cite{Howard1964}, \cite{Tung1981}, and \cite{Lin2003}, to the quasi-geostrophic system and then utilise it to check the parameter dependence of eigenvalues. Section \ref{sec:conclusion} concludes with a summary and discussion of the hurdle theorem concept. }

\section{Formulation of the problem}
\label{sec:formulation}

\textcolor{black}{For readers not familiar with geophysics, we introduce some basic terminology before presenting our model.}
Our starting point is quasi-geostrophic conservation of potential vorticity on a beta-plane \citep{Vallis2017},
$\frac{DQ}{Dt}=Q_t-\Psi_y Q_x+\Psi_x Q_y=0$, 
where $\Psi$ is the streamfunction for the predominantly horizontal flow and the quasi-geostrophic potential vorticity is written as
$Q=f_0+\beta y+\Psi_{xx}+\Psi_{yy}+\frac{f_0^2}{\rho}\left ( \frac{\rho}{N^2}\Psi_z \right )_z$.
Here, $t$ is time and $x$, $y$, and $z$ are spatial coordinates in the zonal (longitudinal), meridional (latitudinal), and vertical directions, respectively; when these appear as a subscript, the meaning is partial differentiation. The zonal and meridional wind components can be found as $U = -\Psi_y$ and $V = \Psi_x$, respectively. In the quasi-geostrophic framework, the static density and squared buoyancy frequency, $\rho$ and $N^2$, are given functions of $z$.

The quasi-geostrophic equation can be derived by first applying the shallow \textcolor{black}{layer} approximation to the Navier-Stokes equations and then taking the limit of strong stratification and rapid rotation. Conversely, the KA-I stability condition can be extended from the quasi-geostrophic framework to the primitive shallow-water framework \citep{Ripa1983}. 
\textcolor{black}{In the primitive context, note that admission of buoyancy (gravity) waves in addition to Rossby waves complicates the shear stability question; however, the quasi-geostrophic results continue to be useful \cite[e.g.,][figures 3 and 5]{Stamp1993}.}

In the above formulation, the terms $f_0+\beta y$ are the first two terms of the Taylor series of the planetary vorticity, $f$, which is also called the Coriolis parameter. The terms $\Psi_{xx}+\Psi_{yy}$ are the relative vorticity. The last term involving two vertical derivatives is the stretching vorticity. If separation of variables can be used in $z$, the stretching vorticity may be simplified via the vertical eigenvalue problem
\begin{eqnarray}
\frac{f_0^2}{\rho}\left ( \frac{\rho}{N^2}\varPhi_z \right )_z=-L_d^{-2}\varPhi \, ,
\end{eqnarray}
with suitable boundary conditions, in which case there is a different Rossby deformation length, $L_d$, for each vertical mode. Usually it is the smallest $L_d^{-2}>0$ (the first baroclinic mode) that is of interest. 
The corresponding first Rossby deformation length (hereafter simply denoted as $L_d$)
 is the length scale that separates potential-energy dominated structures, i.e., large-scale pressure highs and lows maintained by the Coriolis effect, from kinetic-energy dominated structures, i.e., small-scale pressure anomalies that get flattened by gravity. 
The physical role played by $L_d$ is analogous to the Larmor radius in plasma flows \citep{Hasegawa1985}.


\subsection{The $1\frac{3}{4}$ model and the eigenvalue problem}

\textcolor{black}{Our formulation is based on multi-layer quasi-geostrophic systems 
\citep[see section 5.3.2 of][]{Vallis2017}. }
It is common practice in geophysical fluid dynamics to simplify the problem by studying a multi-layer shallow-water model in which constant-density layers are arranged in a (hydro)statically stable manner, with low density overlying high density. A two-layer model of this type, with a fully dynamic weather layer that overlies a layer containing a deep jet profile, $U_{\rm deep}$, was first applied to Jupiter by \cite{Ingersoll1981}. 
In this case, \textcolor{black}{the weather layer potential vorticity is written as}
\begin{eqnarray}
Q=f_0+\beta y+\Psi_{xx}+\Psi_{yy}-L_d^{-2}(\Psi-\Psi_{\rm deep})
\label{QGdp} \, ,
\end{eqnarray}
where $L_d$ is the first Rossby deformation length 
and $\Psi_{\rm deep}$ is the deep-layer streamfunction, also known as the dynamic topography. 
\textcolor{black}{
The ratio of the depth of the deep layer to the weather layer is taken to be very large, so that $\Psi_{\rm deep}$ may be treated as steady \citep{Majda2005}.}
\textcolor{black}{We further assume for simplicity that the deep-layer circulation does not vary with longitude, $x$. Such variations can affect the phase-locking of long Rossby waves and thereby can play an indirect role, however the focus here is on unidirectional shear instability, which is relevant given the predominantly zonal nature of gas giant circulations. Under those assumptions,} $\Psi_{\rm deep}$ and $U_{\rm deep} = -(\Psi_{\rm deep})_y$ are functions of $y$ only. 
The weather layer corresponds to the first-baroclinic-mode structure of the atmosphere of a gas giant, while the barotropic structure of the gas-giant interior is modelled by $\Psi_{\rm deep}$. This two-layer configuration has traditionally been called the ``\textcolor{black}{$1\frac{1}{2}$} layer'' model, but was recently re-branded as the ``\textcolor{black}{$1\frac{3}{4}$} layer'' model when $U_{\rm deep}$ has an alternating jet profile rather than being still ($U_{\rm deep} = 0$), to emphasise the effect on the weather-layer dynamics that the corresponding undulations of $\Psi_{\rm deep}$ have via stretching vorticity \citep{Flierl2019}.

\textcolor{black}{In summary, the $1\frac{3}{4}$ layer model is written as conservation of quasi-geostrophic potential vorticity \textcolor{black}{$\frac{DQ}{Dt}=0$} for (\ref{QGdp}) which now yields one nonlinear} equation in one unknown, $\Psi(x,y,t)$. \textcolor{black}{As previously noted,} gas giants are observed to be predominantly zonal, hence for the linear stability analysis the basic state is assumed to be only a function of latitude, $y$. The $x$ dimension on a gas giant is periodic, hence the $x$ and $t$ dependencies are represented with a Fourier component by replacing $\Psi(x,y,t)$ in (\ref{QGdp}) with $\Psi(y)+\delta \, \psi(y) \exp[ik(x-ct)]$, where $k$ and $c$ are the zonal wavenumber and phase speed, respectively, and $\delta$ sets the scale of the amplitude. The $y$ dimension is assumed to span a channel of width $L$ centred on the origin, $y \in (-L/2,L/2)$. When considering Jupiter's atmosphere, we assume that the channel is contained inside the northern or southern extratropical domain (i.e., poleward of the equatorial jet), where quasi-geostrophic theory holds. The meridional boundary conditions on the perturbations are Dirichlet, $\psi(-L/2)= 0$ and $\psi(L/2)=0$, unless otherwise stated. 

The problem is linearised by restricting to $|\delta| \ll 1$ and retaining only $O(\delta)$ terms.  These standard assumptions yield a linear ordinary differential equation that governs the meridional structure of small-amplitude perturbations, 
\begin{equation}
\psi''-(k^2+L_d^{-2})\psi +\frac{Q'}{U-c}\psi=0 \, , \qquad y \in \Omega \, ,
\label{EQ}
\end{equation}
where $\Omega=(-L/2,L/2)$ and a prime denotes ordinary differentiation with respect to $y$. 
Three length scales exist for this problem: the channel width, $L$, the deformation length, $L_d$, and the scale at which the basic flow varies, $L_U$.  In what follows, dimensional expressions are retained when addressing physical phenomena, but in the mathematical case studies expressions are implicitly non-dimensionalised \textcolor{black}{using the length scale $L_U$ (i.e., $L$ and $L_d$ are $L/L_U $ and $L_d/L_U$ in the dimensional form, respectively).}

In the \textcolor{black}{$1\frac{3}{4}$} layer model, $U(y)=-\Psi'(y)$ and $Q'(y)$ are linked by
\begin{eqnarray}
Q'(y)=\beta -U''(y)+L_d^{-2} \big( U(y)-U_{\rm deep}(y) \big) \, .
\label{Q'}
\end{eqnarray}
\textcolor{black}{Note that in this model a Galilean transformation in $x$ is applied for both layers so that $U$, $U_{\rm deep}$ are replaced by $U-\alpha$, $U_{\rm deep}-\alpha$, respectively, implying that $Q'$ is unchanged (note that this is not the case in the $1\frac{1}{2}$ layer model, where $U_{\rm deep}$ is related to bottom topography and hence not altered in the Galilean shift).}
The link (\ref{Q'}) implies that two of the three basic flow profiles, $U(y)=-\Psi'(y)$, $U_{\rm deep}(y)$, and $Q'(y)$
may be specified independently, after which the third is fixed. \textcolor{black}{This is important in planetary science, as} observing $U$ and $Q'$ determines $U_{\rm deep}$, providing new insights into the nature of deep jets \citep{Dowling1995_estimate,Read2006,Read2009,Read2009b}. 
%
\textcolor{black}{Furthermore, as we shall see shortly, if there is a neutrality hypothesis that can potentially be used to constraint the reciprocal Rossby Mach number, to be defined in (\ref{RosM}), then $U$ and $Q'$ can be related.} 
This is useful as $U$ is easily observable, while precise measurements of $Q'$ require accurate temperature retrievals and thus pose a relatively greater challenge. Physically, the gas giant zonal winds are observed to be quite \textcolor{black}{stable;} hence in planetary physics, determining neutral Rossby Mach number profiles has been a focus.

Given a wavenumber, $k\geq 0$, and deformation radius, $L_d> 0$, equation (\ref{EQ}) and the boundary \textcolor{black}{conditions constitute} an eigenvalue problem for the complex phase speed, $c=c_r+ic_i$. For fixed $L_d$, if there is a $k$ that yields \textcolor{black}{$c_i> 0$, the flow is unstable. 
It is easy to confirm that the complex conjugate of the eigenvalue is also an eigenvalue; however, when $|c_i|$ is small, only the mode with $c_i>0$ provides a good approximation to the viscous problem.}

\textcolor{black}{In the wider context, it is important to note that our mathematical analysis holds for the quasi-geostrophic equation linearised around a broad range of $U(y)$ and $Q'(y)$ profiles.
In the case of rotation with a non-zero planetary vorticity gradient, $\beta \neq 0$, the necessary and sufficient conditions for stability have also been discussed in the context of the analogous problem in plasma physics \citep[e.g.,][]{Numata_etal2007,Zhu_etal2018}. In the case of no rotation and no stratification, $\beta = 0$ and $L_d \rightarrow \infty$, the PV gradient (\ref{Q'}) simplifies to $Q' = -U''$ and (\ref{EQ}) reduces to the original Rayleigh equation, in which case the critical latitudes are inflection points of $U(y)$. }



%
\section{The stability conditions}

\textcolor{black}{The stability conditions for the eigenvalue problem can be succinctly expressed using the reciprocal of the Rossby Mach number, as we shall see shortly.}
In the subsequent two subsections, we define two classes of basic flows: Class (i) and Class (ii).
This classification is important as the strength of the stability conditions varies between them.
It is convenient to define the set of the \textcolor{black}{PV extrema} $Y_Q=\{y \in \Omega | Q'(y)=0\}$ and the set of the zonal wind \textcolor{black}{zeros} in a suitable Galilean frame $Y_{U,\alpha}=\{y \in \Omega | U(y)=\alpha\}$; the number of elements of the latter set depends on $\alpha$. 
Hereafter we assume that $Q'(y)$, $U(y)$ are $C^1$.
Therefore, unless $(U-c)$ vanishes somewhere, the eigenfunctions belong to the function space
\begin{eqnarray}
C_0^2=\{f\in C^2(\overline{\Omega})| f(-L/2)=f(L/2)=0\},
\label{eq:C02}
\end{eqnarray}
where $\overline{\Omega}=[-L/2,L/2]$. \textcolor{black}{For neutral modes, the critical latitudes are expressed as $Y_{U,c}$.}

\subsection{The reciprocal Rossby Mach number}




\textcolor{black}{
The reciprocal Rossby Mach number, $M^{-1}(y)$ emerges as being central to this work and there are various ways to motivate it. For example, in Section 1 we briefly commented on the physical interpretation by \cite{Dowling2014,Dowling2020}.
The following motivation is based on mathematical consideration applied to the second order ordinary differential equation (\ref{EQ}). 
If equation (\ref{EQ}) possesses a non-trivial solution that meets the boundary conditions, then in order for the solution to exhibit oscillatory behaviour at some point, the factor $Q'/(U-c)$, needs to be sufficiently larger than the smallest eigenvalue, $\kappa_0^2$, of the eigenvalue problem formulated by the remaining terms with $k=0$:}
\begin{eqnarray}
\varphi''-L_d^{-2}\varphi=-\kappa^2 \varphi, \qquad y \in \Omega \label{phiEQ}
\end{eqnarray}
with $\varphi(-L/2)=\varphi(L/2)=0$. 
This leads to the introduction of the reciprocal Rossby Mach number,
\begin{eqnarray}
M_{\alpha}^{-1}(y)=\frac{1}{\kappa_0^2}\frac{Q'}{U-\alpha} \, ,\label{RosM}
\end{eqnarray}
where the subscript alpha explicitly shows the dependence on a reference-frame shift of the zonal wind, $\alpha \in \RealN$. 
\textcolor{black}{At this stage, $\alpha$ is arbitrary. However, as we will explain later, in the Jupiter-style basic flows, there is only one optimal choice of $\alpha$, allowing us to remove the subscript.}
Note from (\ref{Q'}) that $Q'$ is invariant with respect to $\alpha$. Both $\kappa_0^2$ and its associated eigenfunction, $\varphi_0$, can be explicitly computed:
\begin{eqnarray}
\kappa_0^2=\frac{\pi^2}{L^2}+L_d^{-2},\qquad \varphi_0=\cos\left (\frac{\pi}{L}y \right).\label{kappa0}
\end{eqnarray}

\subsection{A sufficient condition for stability: Class (i) basic flows}
The extant sufficient conditions for stability can be expressed in terms of $M_{\alpha}^{-1}$ as follows.
\begin{thm}\label{KA}
Suppose there exists $\alpha$ such that $M_{\alpha}^{-1}\leq 0$ for all $y\in \Omega$, or that $0\leq M_{\alpha}^{-1} \leq 1$ for all $y\in \Omega$. Then there is no unstable mode for any $k$.
\end{thm}
\noindent
Here, the former and latter conditions correspond to KA-I and KA-II, respectively. Mathematically, the stability conditions can be used in the equality cases, although they are often omitted in the physics literature \citep{Dowling2014}.
\textcolor{black}{Our system is classified as a non-canonical Hamiltonian system, and the theorem can be shown by Arnol'd's method; if a Hamiltonian $H$ and a Poisson bracket $\{,\}$ satisfy $\delta H:=\{F,H\}=0$ for all functionals $F$ at a steady solution, then the solution is stable if $\delta^2 H:=\{F,\{F,H\}\}$ is strictly positive or negative definite for all $F$.}
The method has a wide range of applications and can also derive nonlinear stability with respect to finite amplitude disturbances. However, if we restrict ourselves to the linear eigenvalue problem, Theorem \ref{KA} can be proved by a more straightforward approach, as shown in Appendix A.

\begin{figure}
\begin{center}
  \includegraphics[width=0.9\textwidth]{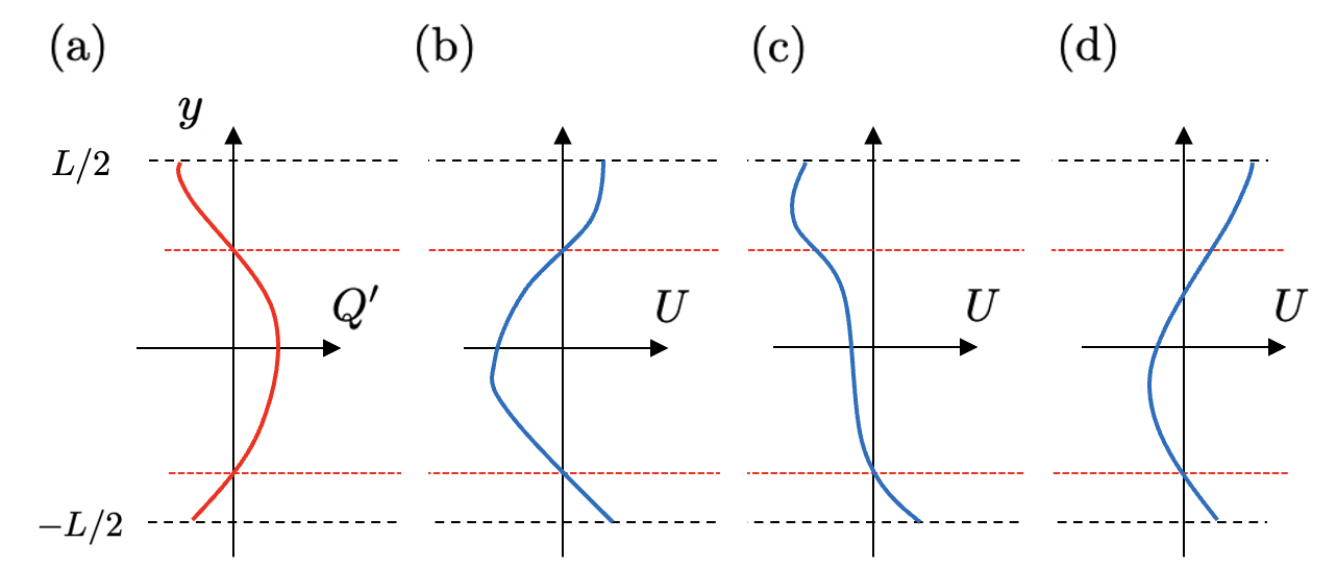}
\end{center}
\caption{Classification of basic flow profiles. For all these examples, consider the PV gradient $Q'(y)$ profile shown in panel (a). The two \textcolor{black}{PV extrema} are indicated by the horizontal red dashed lines. 
The $U(y)$ profile shown in panel (b) is Class (i). 
The $U(y)$ profile shown in panel (c) is Class (ii) but not Class (i).
The $U(y)$ profile shown in panel (d) is not Class (ii).  
}
\label{fig:class}
\end{figure}
If $Q'(y)$ does not change sign over $\Omega$, \textcolor{black}{i.e. there are no PV extrema,} we can always choose a large enough or small enough $\alpha$ to satisfy KA-I, which recovers the Charney-Stern condition 
\citep[KA-I is also known as a sufficient condition for the absence of over-reflections of Rossby waves; see][]{LindzenTung1978}. 
In other words, the interesting case occurs when there is at least one \textcolor{black}{PV extremum}.
We assume the following properties for both Class (i) and Class (ii) basic flows:
\smallskip
\begin{itemize}
\item The PV gradient $Q'(y)$ changes sign somewhere in $\Omega$. 

\item The zeros of $Q'(y)$, the \textcolor{black}{PV extrema, are isolated}. 

\item $Q''(y_l)\neq 0$ for all $y_l \in Y_Q=\{y \in \Omega | Q'(y)=0\}$.
\end{itemize}
\smallskip

We say a basic flow belongs to Class (i) if there exists an $\alpha \in \mathcal{R}$ such that the following additional condition is satisfied.
\begin{itemize}
\item $M_{\alpha}^{-1}(y)$ is continuous and one-signed \textcolor{black}{in $\Omega$.}
\end{itemize}
Here $\mathcal{R}$ is the range of the zonal flow
\begin{eqnarray}\label{mathscrR}
\mathcal{R}=\left(\min_{y\in \overline{\Omega}}U,\max_{y\in \overline{\Omega}} U\right ).
\end{eqnarray}
In the definition, `a function is one-signed' means that it is non-negative or non-positive for all $y \in \Omega$. \textcolor{black}{Of course, from Theorem 1, the non-positive cases are stable.}

\begin{figure}
\begin{center}
  \includegraphics[width=0.7\textwidth]{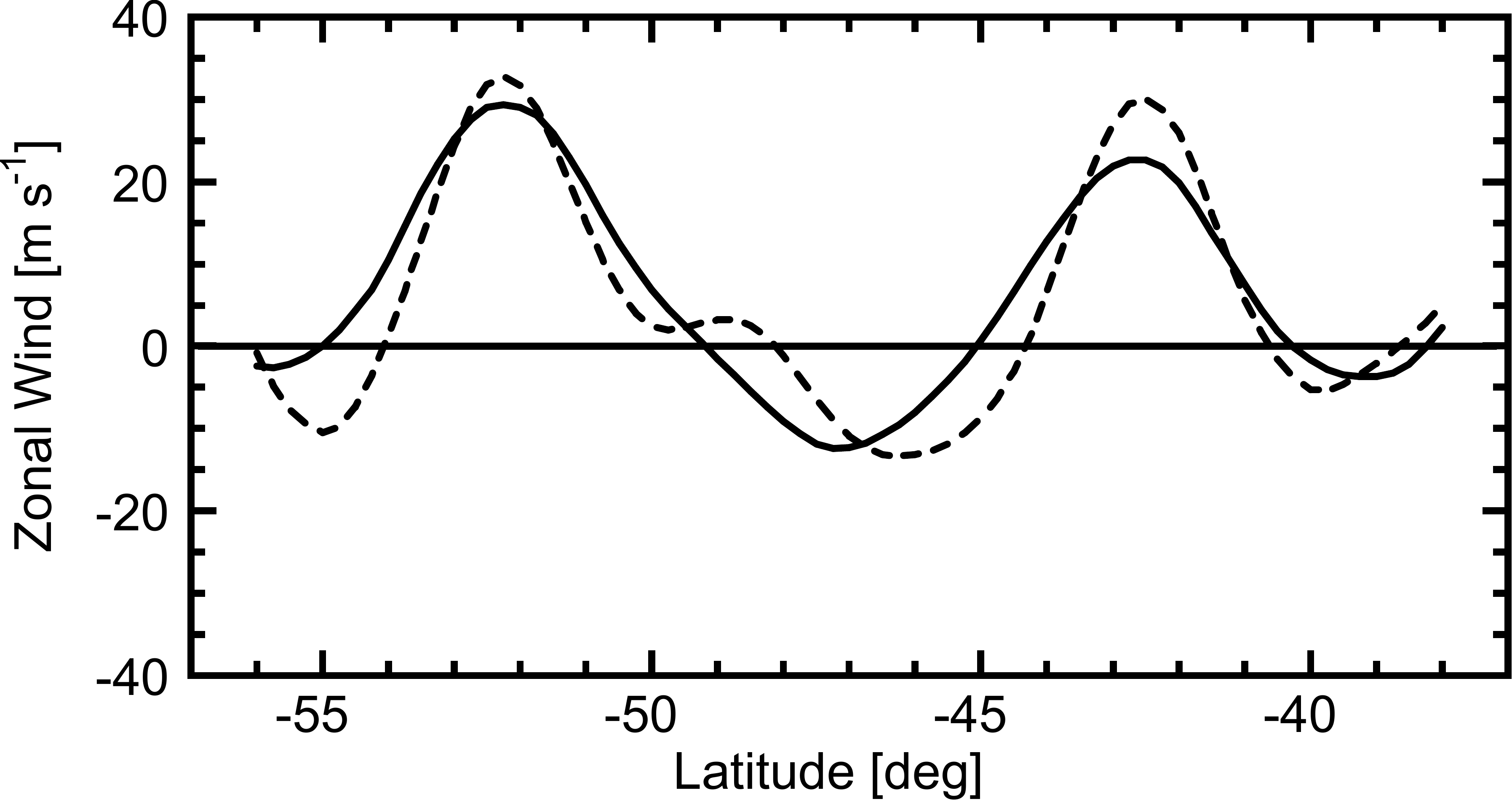}
\end{center}
\caption{
\textcolor{black}{Comparison of Jupiter profiles of $U$ (solid) and $\kappa_0^{-2}Q_y$ (dashed), using the same data as in figure \ref{fig:Read2006_profiles}. 
Since $L/L_d \gg 1$ on Jupiter, it follows that $\kappa_0^2 \approx L_d^{-2}$, such that the similarity between the two profiles implies $M^{-1} \approx 1$ with $\alpha \approx 0$. This, and further evidence discussed in the text, support the identification of Class (i) as the Jupiter-style class.
The deformation length is set here to $L_d = c_{\text{gw}}/|f_0| = 1750 \, \rm km$, using the gravity wave speed, $c_{\text{gw}} = 454 \, \rm m \, s^{-1}$, inferred from the Comet Shoemaker-Levy 9 impact analysis \citep{Hammel_1994} and the Coriolis parameter, $f_0$, evaluated at the G-fragment impact site, latitude $\phi_0 = -47.5^\circ$ (planetographic); the figure is centred on $\phi_0$. }
}
\label{fig:Read2006_profiles2}
\end{figure}
Class (i) occurs when the \textcolor{black}{PV extrema} and the \textcolor{black}{zeros} of $U-\alpha$ coincide, i.e. $Y_Q=Y_{U,\alpha}$.
For example, this is the case when $Q'$ in figure \ref{fig:class}a is paired with $U$ in figure \ref{fig:class}b. 
On the other hand, the basic flow is not Class (i) when $U$ is replaced by the profile shown in figure \ref{fig:class}c or d.

The importance of Class (i) can be seen from the fact that it is the case that Theorem 1 may be able to show stability, when there is a \textcolor{black}{PV extremum}.
%
%
The KA-I and KA-II stability conditions can be combined; 
the flow is stable if 
\begin{eqnarray}
\max_{y\in \overline{\Omega}} M^{-1}\leq 1. \label{C1KA}
\end{eqnarray}
Here the reciprocal Rossby Mach number is simply written as $M^{-1}$, as the choice of $\alpha$ is trivial (see Theorem 3 in Section 6.1). \textcolor{black}{Note if $M_{\alpha}^{-1}(y)$ changes sign (i.e. the basic flow is not Class (i)), mathematically the condition (\ref{C1KA}) cannot guarantee stability.} 

Class (i) is the geophysically interesting case to which the ``Jupiter-style'' shear flow belongs, \textcolor{black}{assuming a suitable $U_{\rm{deep}}$} \citep{Dowling2020,Afanasyev2022}. 
As mentioned in section 1, observations support that $M^{-1}$ is around unity in Jupiter's and Saturn's atmosphere \citep{Dowling1993,Read2006,Read2009,Read2009b}. 
This configuration can be inferred from the most accurate available observational data, as depicted in figure \ref{fig:Read2006_profiles2}. The link between Class (i) and Jupiter's atmosphere is further discussed in Section 5.

\subsection{A sufficient condition for instability: Class (ii) basic flows}

Basic flows that are not Class (i) may also be physically important. For example, $M_{\alpha}^{-1}$ changes sign on a seasonal basis in Earth's atmosphere \citep{Du2015}. Our main result, Theorem 2, which we call the hurdle theorem, can show the existence of instability for a wider range of basic flows than Class (i).

We say a basic flow belongs to Class (ii) if there exists an $\alpha \in \mathcal{R}$ satisfying the following conditions, called Class (ii) conditions.
\smallskip
\begin{itemize}
\item $M_{\alpha}^{-1}(y)$ is continuous \textcolor{black}{in $\Omega$.}
\item $M_{\alpha}^{-1}(y_j)$ is one-signed for all $y_j \in Y_{U,\alpha}$.
\end{itemize}
\smallskip
Note that if a basic flow is Class (i), then it is Class (ii).
However, for Class (ii) basic flows, in general there can be more than one $\alpha$ that make $M_{\alpha}^{-1}$ continuous. In figures \ref{fig:class}a,c, for $\alpha=0$ the lower \textcolor{black}{PV extremum} cancels the singularity and $M_{\alpha}^{-1}$ changes sign at the upper \textcolor{black}{PV extremum}.  Alternatively, an appropriately negative $\alpha$ may be chosen such that $U-\alpha$ vanishes at the upper \textcolor{black}{PV extremum}.

\begin{figure}
\begin{center}
     \includegraphics[width=0.9\textwidth]{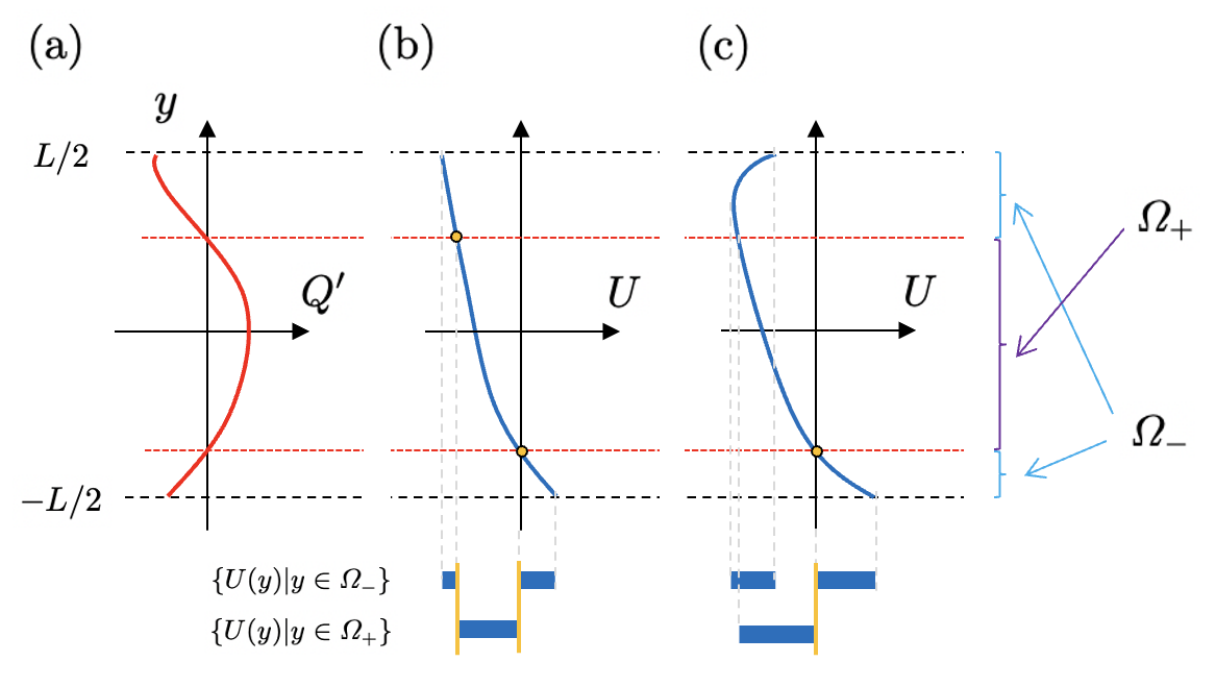}
\end{center}
\caption{The PV gradient profile $Q'(y)$ defines the sets $\Omega_+=\{y\in \Omega| Q'(y)>0 \}$, $\Omega_-=\{y\in \Omega| Q'(y)<0 \}$. For the zonal wind profiles $U(y)$ shown in panels (b) and (c),  
the set $\mathcal{D}=\mathcal{R}\backslash (\{U(y)|y\in \Omega_+\}\cup \{U(y)|y\in \Omega_-\})$, has two and one points, respectively \textcolor{black}{(indicated by orange bullets and lines)}. 
Furthermore, in the vicinity of these points, there are no points belonging to the set (\ref{union}).
Therefore both cases are Class (ii). 
}
\label{fig:disjoint}
\end{figure}
It is useful to define the sets $\{U(y)|y\in \Omega_+\}$ and $\{U(y)|y\in \Omega_-\}$, decomposing $\Omega$ into the three parts, $\Omega_+=\{y\in \Omega| Q'(y)>0 \}$, $\Omega_-=\{y\in \Omega| Q'(y)<0 \}$, and $Y_Q$. 
Then if the set
\begin{eqnarray}
\mathcal{D}=\mathcal{R}\backslash (\{U(y)|y\in \Omega_+\}\cup \{U(y)|y\in \Omega_-\})
\end{eqnarray}
is non-empty, we may be able to select $\alpha$ from this set. 
\textcolor{black}{Furthermore, in the vicinity of the chosen point, there should be no points belonging to 
\begin{eqnarray}
\{U(y)|y\in \Omega_+\}\cap \{U(y)|y\in \Omega_-\}\label{union}
\end{eqnarray}
to satisfy the second Class (ii) condition. }

To visually check whether the basic flow is Class (ii), first find $\Omega_+$ and $\Omega_-$ using $Q'(y)$ and then illustrate $\{U(y)|y\in \Omega_+\}$ and $\{U(y)|y\in \Omega_-\}$.
For example, with $U(y)$ shown in figure \ref{fig:disjoint}b and c, the number of elements in $\mathcal{D}$ is one and two, respectively. The condition that the point at which $\{U(y)|y\in \Omega_+\}$ and $\{U(y)|y\in \Omega_-\}$ are separated in the range of $U(y)$ is not particularly restrictive, such that a wide array of basic flows belongs to Class (ii). In particular, if $U(y)$ is monotonic and there is at least one \textcolor{black}{PV extremum}, $\mathcal{D}$ should be non-empty. On the other hand, for the basic flow $U(y)$ shown in figure \ref{fig:class}d, $\mathcal{D}$ is empty, and in fact the basic flow cannot be Class (ii).

Our main result is the following `hurdle theorem', which provides a sufficient condition for instability (and the contrapositive necessary condition for stability). This theorem is proved in Section 6:
\begin{thm}\label{ThmM}
Suppose the basic flow is Class (ii). Fix an $\alpha$ so that the Class (ii) conditions are satisfied.
The flow is unstable if there is an interval $[y_1,y_2] \subset \Omega$ over which
\begin{eqnarray}
h\equiv \frac{\frac{\pi^2}{(y_2-y_1)^2}+L_d^{-2}}{\frac{\pi^2}{L^2}+L_d^{-2}}<  M_{\alpha}^{-1}\label{hhh}
\end{eqnarray}
is satisfied.
\end{thm}
\begin{figure}
\begin{center}
  \includegraphics[width=0.9\textwidth]{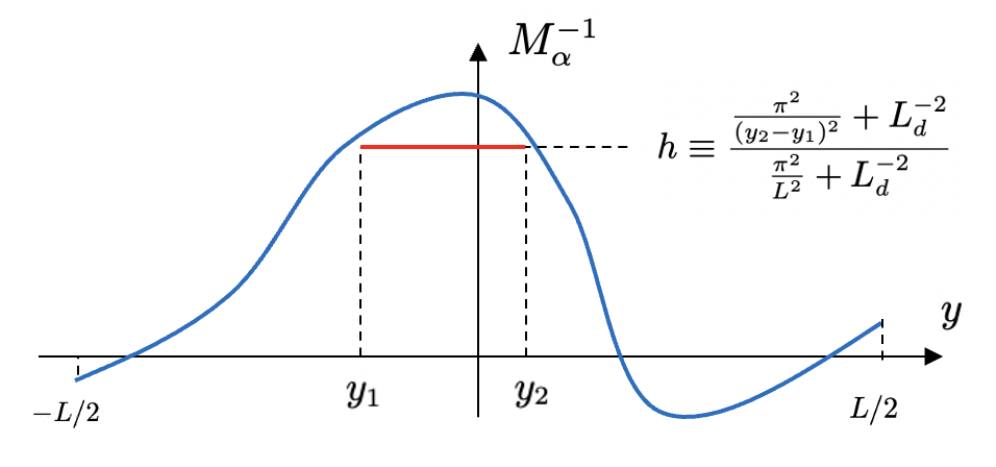}
\end{center}
\caption{An example of unstable $M^{-1}_{\alpha}(y)$ profile. Observe that the blue curve overcomes the hurdle $h$.
}
\label{mcompare}
\end{figure}

In short, if the reciprocal Rossby Mach number profile surpasses the hurdle $h$, then the flow is unstable, as illustrated in figure \ref{mcompare}. As can be seen from the proof in Section 6, this result is actually 
\textcolor{black}{independent of the conditions applied at the boundaries (though it depends on $L$).}

Note that the height of the hurdle, $h$, is always higher than unity, and depends on the width of the hurdle. However, if $L_d^{-2} \gg \pi^2/(y_2-y_1)^2$,
the hurdle height is almost 1. Thus, in this limit the condition for the existence of instability asymptotes to
\begin{eqnarray}
\max_{y\in \overline{\Omega}} M_{\alpha}^{-1}>1.\label{CKA}
\end{eqnarray}
This is the contrapositive of (\ref{C1KA}), implying that the KA-II stability condition is almost sharp if the basic flow is Class (i) and $L_d$ is small \textcolor{black}{(recall that for Class (ii) the stability condition (\ref{C1KA}) is not valid).}

\section{Analysis of model profiles}
\label{sec:models}
In this section, we stress-test the hurdle theorem (Theorem 2) with model basic flow profiles. 
\cite{Stamp1993} used the assumption $Q'=aL_d^{-2}U$, such that $M^{-1}_0$ becomes a constant $a$, together with a sinusoidal zonal wind profile $U =\cos(2\pi y/L)$, and applied periodicity in both meridional and zonal directions. 
The assertion of Theorem 1 remains the same with Neumann or periodic boundary 
conditions for (\ref{phiEQ}), in which case (\ref{kappa0}) is replaced by
\begin{eqnarray}
\kappa_0^2=L_d^{-2},\qquad \varphi_0=1 \, . 
\label{kappa0b}
\end{eqnarray}
Their basic flow is clearly Class (i), and the numerical results are consistent with this theoretical result. 
Moreover, their numerical computations suggest that whenever $M_0^{-1}>1$, unstable modes exist for some $k$, implying that KA-II may give a sharp stability boundary. Theorem 2 validates this numerical result, because when $M_0^{-1}$ is a constant, we are able to employ the full-width hurdle with unit height, $h=1$.


\textcolor{black}{In general, the reciprocal Rossby Mach number will not be a constant but rather a profile that may be \textcolor{black}{``supersonic''} in some regions, $M^{-1}_\alpha(y) < 1$,
and \textcolor{black}{``subsonic''} in others, $M^{-1}_\alpha(y) > 1$. To explore the ramifications of this generality to shear instability, we next develop a model with a two-parameter, variable $M^{-1}_\alpha(y)$ profile.  We use this model to numerically explore first Class (i) profiles in section \ref{sec:models}.1, and then Class (ii) profiles that are not Class (i) in section \ref{sec:models}.2. 
Section 5 delves into the connection between the model and the stability of Jupiter's alternating jet streams.
\textcolor{black}{As commented at the end of Section 2, our problem covers the classical Rayleigh equation problem. }
In Appendix C we analyse one more model flow to clarify where our analysis stands in the long history of Rayleigh-equation research. 
}



\subsection{Class (i)}
\begin{figure}
\hspace{20mm}(a)\hspace{55mm}(b)\\
\begin{center}
  \includegraphics[width=0.4\textwidth]{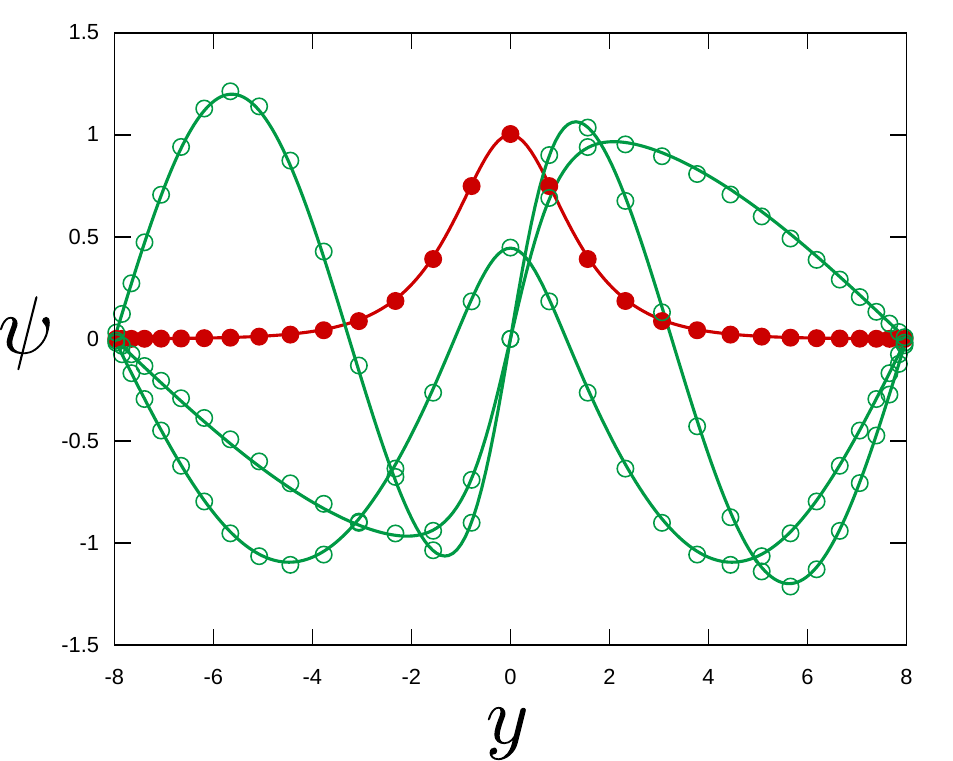}
    \includegraphics[width=0.4\textwidth]{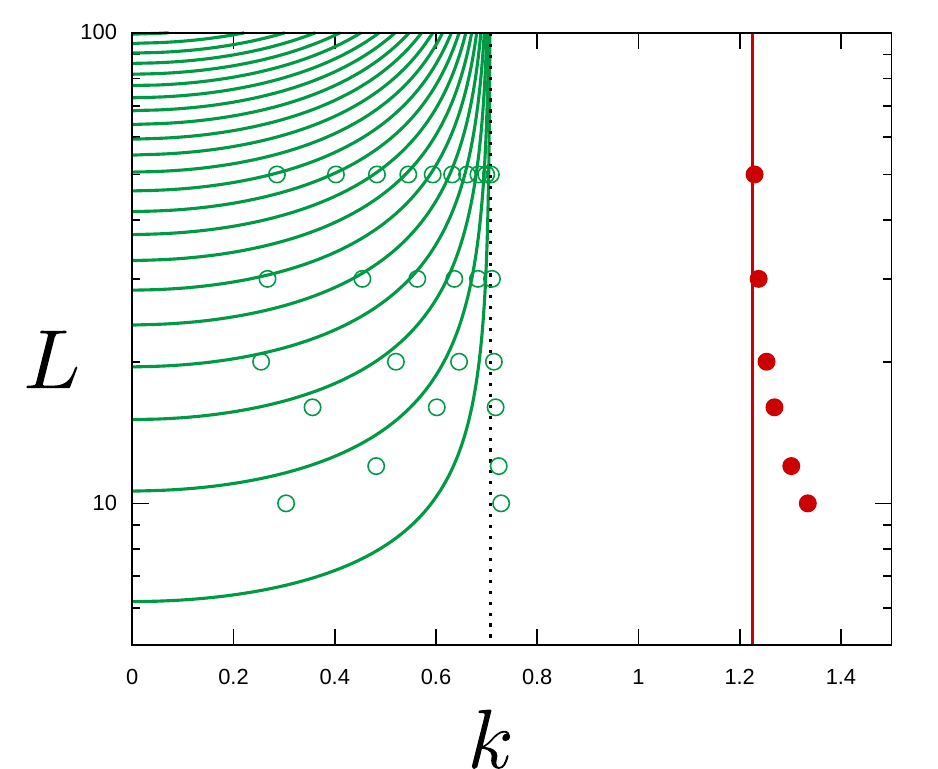}
\end{center}
\caption{
(a) The eigenmodes of (\ref{nEQ}) for $L=16, L_d=1, a=1.5, b=3.5$. There are four neutral modes for this parameter set. The solid curves are the theoretical results assuming large $L$ ($k\approx 1.225, 0.6707, 0.5494, 0.2476$). The symbols are the numerical results ($k\approx 1.269, 0.7178, 0.6016, 0.3564$). The red curve and the filled circles 
\textcolor{black}{is the localised mode.}
The green curves and the open circles are the oscillatory modes. 
(b) The neutral $k$ for $L_d=1, a=1.5, b=3.5$. The solid curves are the theoretical results, and the larger $L$, the better the agreement with the numerical results, shown by the symbols. In the limit $L\rightarrow \infty$, the oscillatory modes (green curves) form a continuous spectrum in the interval $k\in (0,\sqrt{0.5})$. There is only one localised mode (red curve) at $k=\sqrt{1.5}$.
}
\label{fig5}
\end{figure}

\begin{figure}
\hspace{20mm}(a)\hspace{55mm}(b)\\
\begin{center}
  \includegraphics[width=0.4\textwidth]{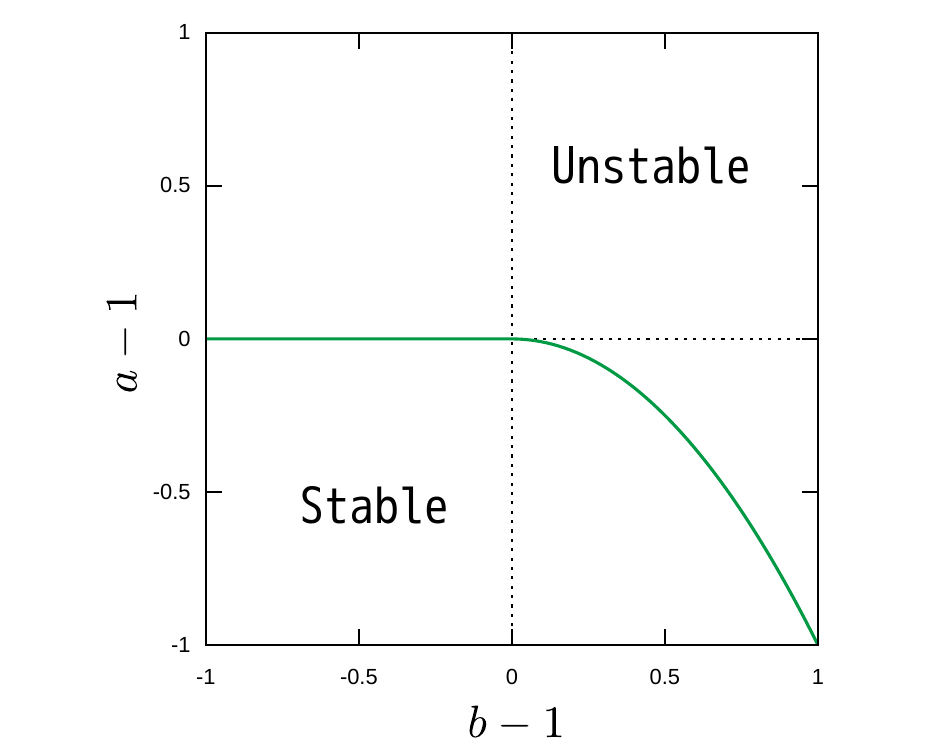}
    \includegraphics[width=0.4\textwidth]{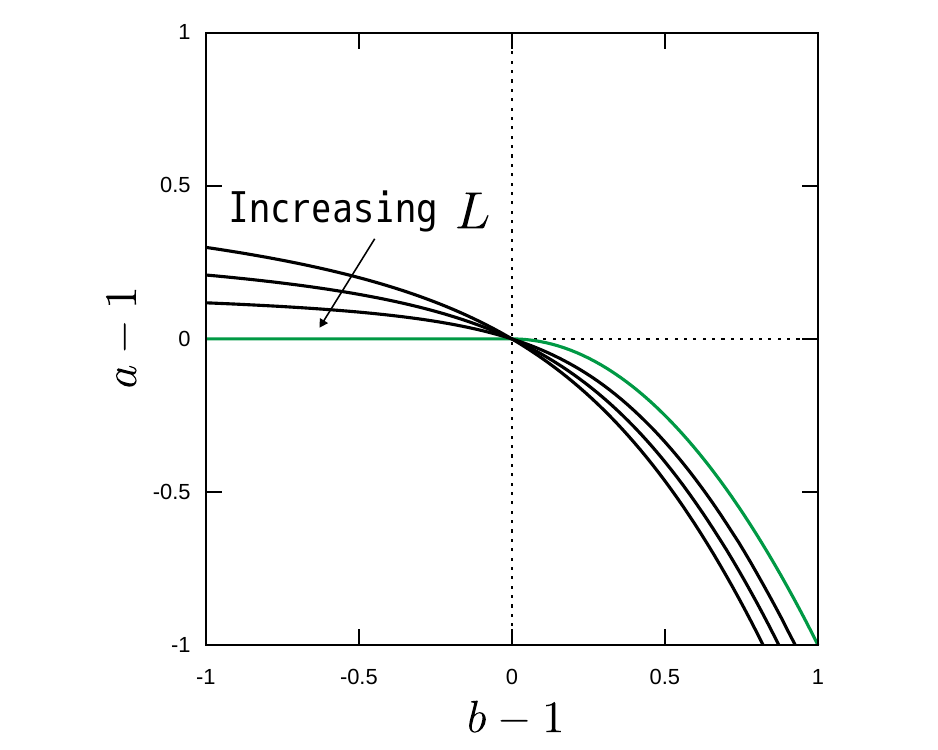}
\end{center}
\hspace{20mm}(c)\hspace{55mm}(d)\\
\begin{center}
      \includegraphics[width=0.4\textwidth]{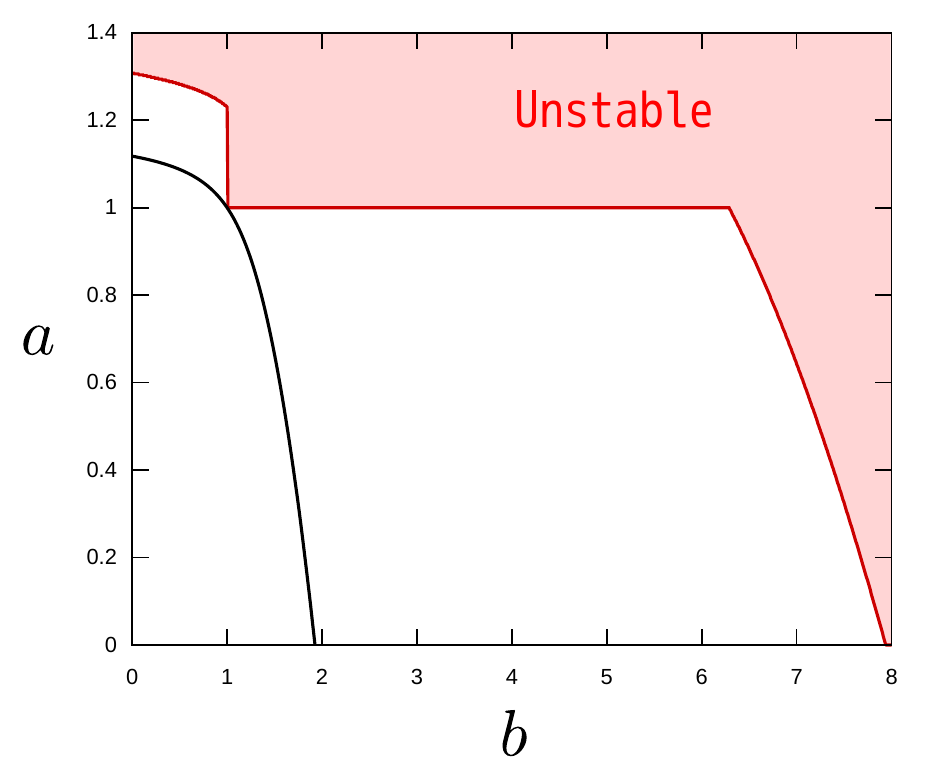}
    \includegraphics[width=0.4\textwidth]{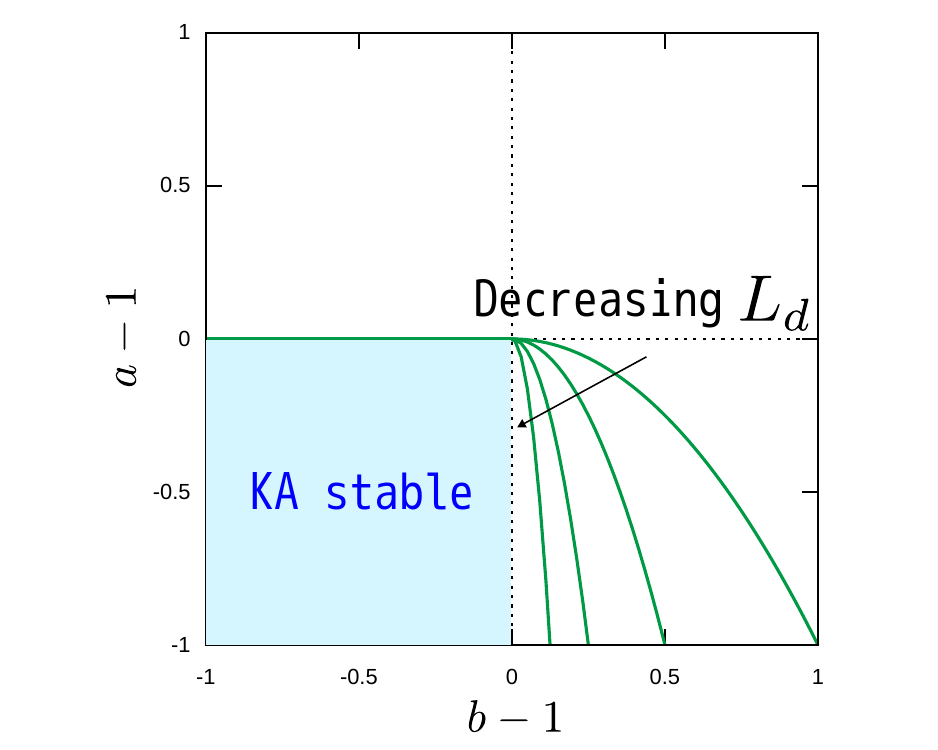}
\end{center}
\caption{(a) The theoretical neutral curve (\ref{Tneutral}) for $L_d=1$.
(b) Comparison of the theoretical neutral curve (green) and the numerical results (black) for $L_d=1$. The three numerical neutral curves are computed using $L=10,12,16$.
(c) Comparison of the numerical neutral curve (black) and the unstable parameter region by Theorem \ref{ThmM} (red shaded). $L=16, L_d=1$. 
\textcolor{black}{The $b$--$a$ plane is used to make it easier to see the relationship with figure 10.}
(d) The theoretical curve (\ref{Tneutral}) for $L_d^{-1}=1,2,4,8$. The blue shaded region is stable from Theorem 1.}
\label{fig6}
\end{figure}

Consider the quasi-geostrophic problem (\ref{EQ}). In setting up a model basic flow, we assume that the PV gradient profile has the form
\begin{eqnarray}
Q'(y)=\kappa^2_0\{a+(b-a)\, \text{sech}^2 (y)\}U(y) \, ,
\label{Qprof}
\end{eqnarray}
where $a,b> 0$ are two free parameters. 
The advantage of using this designed basic flow is that, regardless of $U(y)$, the reciprocal Rossby Mach number with the choice $\alpha=0$ becomes
\begin{eqnarray}
M_{0}^{-1}(y) =a+(b-a)\, \text{sech}^2 (y) \, .
\label{Mprofile}
\end{eqnarray}
The case $a=b$ corresponds to the PV profile considered in \cite{Stamp1993}. When $b > a$, the $M_0^{-1}$ profile has a hump of height $b$ centred at $y=0$, and likewise a dip when $b < a$. 

If $U(y)$ has a \textcolor{black}{zero} (i.e. $0 \in \mathcal{R}$), the existence of a \textcolor{black}{PV extremum} is guaranteed. Then, since both $a$ and $b$ are positive, the function (\ref{Mprofile}) is one-signed for all $y$, implying that the model basic flow belongs to Class (i), i.e. ``Jupiter-style". 
The two different zonal wind profiles 
\begin{eqnarray}
U(y)=\sin(2\pi y),\label{C1B1}\\
U(y)=\tanh(y), \label{C1B2}
\end{eqnarray}
are considered. Here, lengths are implicitly scaled by $L_U$. 
The profile (\ref{C1B1}) may be regarded as a model for Jupiter's alternating jets (with $L_U$ being the jet peak-to-peak length scale), while (\ref{C1B2}) is a simpler case where there is only one \textcolor{black}{PV extremum}.

For Class (i) the phase speed of the neutral modes is uniquely determined (Section 6, Theorem 3), and for our \textcolor{black}{basic flows} it is $0$. Thus, setting $c=0$ in (\ref{EQ}), the equation satisfied by the neutral solutions can be found as
\begin{eqnarray}
\psi''-(k^2+L_d^{-2}-\kappa_{0}^2M^{-1})\psi = 0 \, ,
\label{nEQ}
\end{eqnarray}
simplifying the notation as $M^{-1}=M_0^{-1}$.
This equation and the Dirichlet boundary condition constitute an eigenvalue problem with $\lambda=-k^2$ as the eigenvalue. Considering that the \textcolor{black}{model flows are meant to emulate} the atmosphere of Jupiter, contemplating the case of large $L$ is natural. 

\begin{figure}
\qquad \qquad (a)\\
\begin{center}
(b) \includegraphics[width=0.8\textwidth]{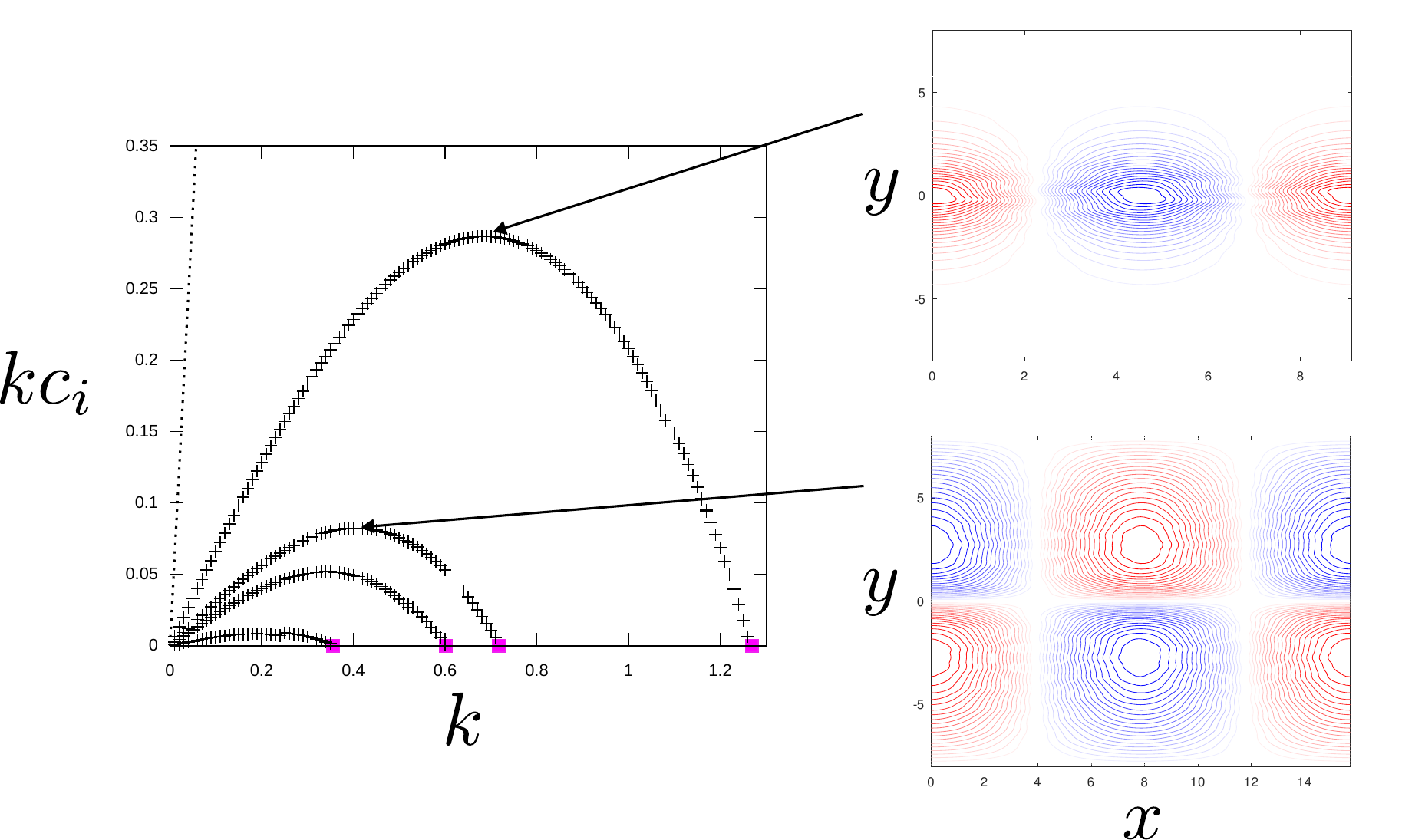}\\
~~~ \includegraphics[width=0.8\textwidth]{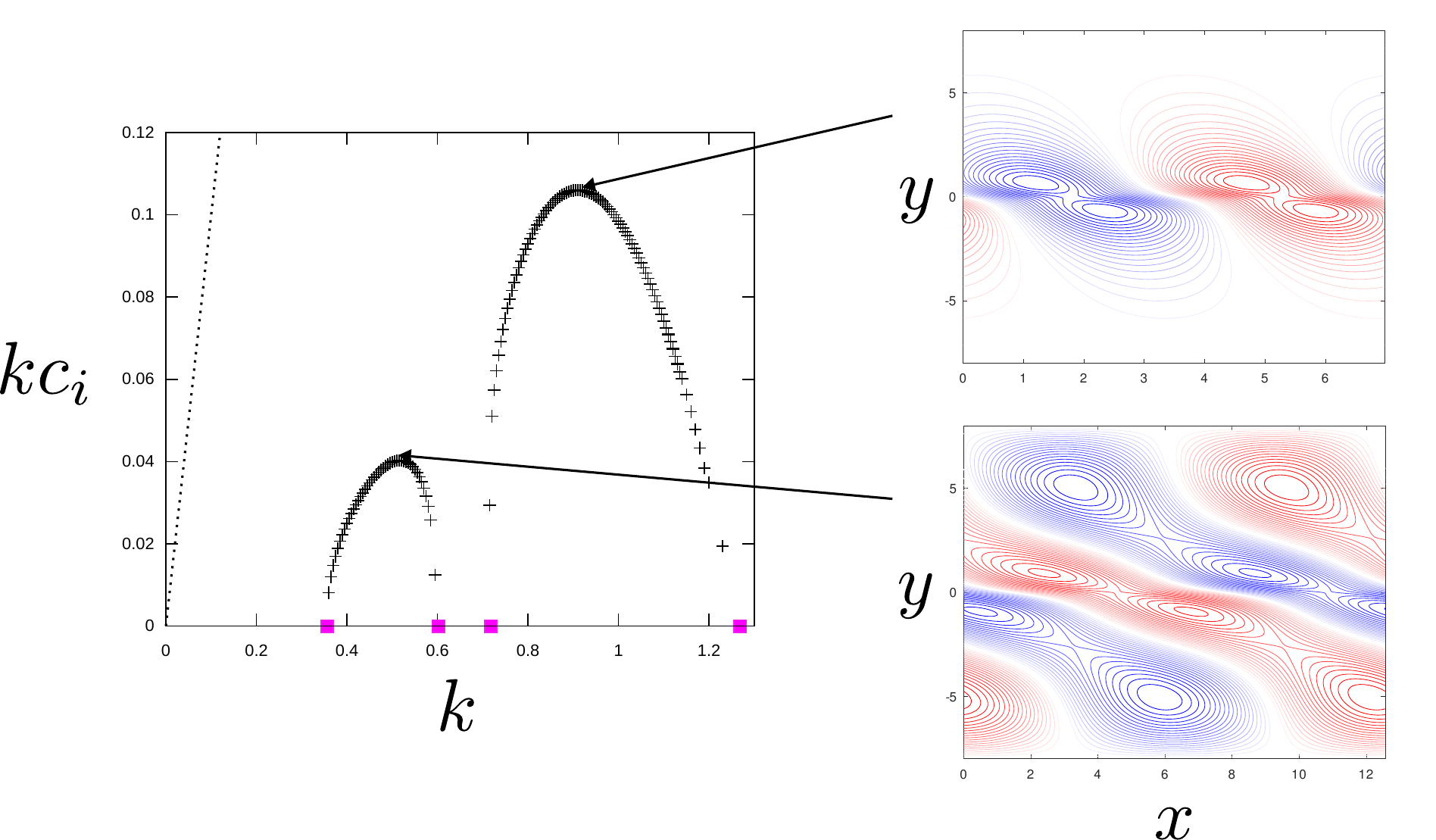}
\end{center}
\caption{The black crosses are numerical solutions of the eigenvalue problem (\ref{EQ}) with the PV gradient profile (\ref{Qprof}) and the parameters 
$L=16, L_d=1, a=1.5, b=3.5$.
The magenta squares are the neutral modes computed in figure \ref{fig5}a. 
The dotted curves indicate the analytic upper bound of $c_i$ \textcolor{black}{shown in the supplementary material.} 
The panels on the right depict contour plots of the streamfunction $\psi(y)e^{ikx}+\psi^*(y)e^{-ikx}$ at the wavenumber that maximises the growth rate $kc_i$. The sign of the contours is distinguished by red and blue.
(a) $U=\sin(2\pi y)$, (b) $U=\tanh(y)$. 
}
\label{fig7}
\end{figure}

When $L$ is large, it is possible to study (\ref{nEQ}) analytically.
To see the behaviour of a typical neutral mode, let us choose the parameter $L_d^{-2}(b-a) = 2$ that makes the analysis simple. First, we note that $\psi=\text{sech}y$ approximately satisfies (\ref{nEQ}) when $k=\sqrt{1+L_d^{-2}(a-1)}$, because $\kappa_0^2\approx L_d^{-2}$. This mode decays exponentially for large $|y|$ as shown by the red curve in figure \ref{fig5}a. Such localised modes do not interact with the boundary, and they are robust to changes in $L$. 
However, there are also oscillatory modes that do not show evanescent behaviour near the boundaries; see the green curves in figure \ref{fig5}a. Those curves can also be found theoretically. Let $\omega=\sqrt{L_d^{-2}(a-1)-k^2}$ be real. Then 
$\psi=\omega \cos(\omega y)-(\text{tanh}y)\sin(\omega y)$ and $\psi=\omega \sin(\omega y)+(\text{tanh}y)\cos(\omega y)$ satisfy (\ref{nEQ}). 
If $L$ is large, the former and latter solutions approximately satisfy the boundary conditions when $\tan(\omega L/2)=\omega$ and $\tan(\omega L/2)=-\omega^{-1}$, respectively. There are many $\omega$ that meet these requirements, and whenever $k=\sqrt{L_d^{-2}(a-1)-\omega^2}$ is real, (\ref{nEQ}) has a neutral mode that oscillates near the boundaries. The number of allowed values of $k$ increases with increasing $L$, as shown in figure \ref{fig5}b, and in the limit of $L\rightarrow \infty$ the eigenvalues of the oscillatory modes form a continuous spectrum occupying $k \in (0,L_d^{-1}\sqrt{a-1})$. 

For $L_d^{-2}(b-a) \neq 2$, the theoretical analysis is a bit more complicated (Appendix B). However, the final results are neat and clean as summarised in figure \ref{fig6}a, 
which shows the parameter plane with $b-1$ and $a-1$ as abscissa and ordinate.
Clearly the third quadrant must be stable as \textcolor{black}{labelled} because of the KA stability condition (\ref{C1KA}). 
Moreover, if $L$ is large, the first and second quadrants are unstable; the easiest way to see this is to note that the right hand side of (\ref{DNC}) is almost $a$, because the integral is largely unaffected by what happens in the hump or dip. The analysis just below (B2) also shows that even if $a$ is slightly above 1, many oscillatory modes will appear when $L$ is large. 
The stability of the flow is in fact non-trivial only in the fourth quadrant of figure \ref{fig6}a, but somewhat surprisingly, the stability in this region can be found analytically (see (B4)). In summary, the theoretical neutral curve for $L\gg 1$ can be described as
\begin{equation}
  a = 
    \begin{cases}
      1 \, ,              & \text{if $b \in (0,1)$}  \\
      1-L_d^{-2}(b-1)^2 \, , & \text{if $b \geq 1 \qquad .$}
    \end{cases}
\label{Tneutral}
\end{equation}
This neutral curve delimits the stable and unstable parameter regions, as shown in figure \ref{fig6}a.
In the fourth quadrant, only localised modes appear, and this is the reason why the neutral curve is so simple. 
Applying the shooting or Chebyshev collocation method to (\ref{nEQ}), we can calculate the neutral curve for finite $L$. The computational results indeed approach the theoretical result (\ref{Tneutral}) as $L$ is increased, as shown in figure \ref{fig6}b. 

One of the novel features of Theorem 2 is that, without resorting to eigenvalue computations, applying a simple hurdle yields a useful estimate of the behaviour of neutral curves. 
In figure \ref{fig6}c, the unstable region is depicted, which can be identified by setting the various hurdles for $M^{-1}(y)$.
The numerically calculated neutral curve should be sandwiched between the KA stable region and the unstable region identified by the hurdle theory. The latter region has a piecewise smooth boundary because the behaviour of hurdles in each quadrant is different. The entire first quadrant is unstable, as can be found by considering the full width hurdle of height unity. In the second quadrant the hurdle giving the best results sits between one of the boundaries and the dip. In the third quadrant, when $b$ is large enough the hump will be able to hurdle over. For this to be seen, the value of $b$ must be at least larger than 6 for $L_d=1$, meaning that the hurdle estimation does not give sharp results. 
However, as noted just below (\ref{CKA}), the smaller $L_d$ becomes, the more accurate is
the hurdle at detecting instability. In line with this expectation, the theoretical neutral curve approaches the KA stability boundary with decreasing $L_d$ (figure \ref{fig6}d).

\textcolor{black}{The} neutral modes do not depend on $U(y)$, but the unstable modes stemming from them do. Figure \ref{fig7} shows the eigenvalue $c$ of (\ref{EQ}) calculated for the same parameters as in figure \ref{fig5}a. For the oscillatory zonal flow (\ref{C1B1}), all instabilities persist down to $k=0$ (figure \ref{fig7}a). However, for the monotonic zonal flow (\ref{C1B2}), the first and second neutral modes, as well as the third and fourth neutral modes, are connected as seen in figure \ref{fig7}b, resulting in a qualitatively completely different diagram. For both cases, the growth rate $kc_i$ is well below the analytically derived upper bound shown by the dotted curve \citep[see supplementary material where Pedlosky’s bound is tightened using the approach by][]{Deguchi2021}.
The streamfunction at the maximum growth rate is indicated by arrows. The fastest growing mode inherits the properties of the localised neutral mode and forms a strong vortex in the centre of the region. The second and subsequent modes spread over the whole domain, like the oscillatory neutral modes.

\begin{figure}
\begin{center}
  \includegraphics[width=0.8\textwidth]{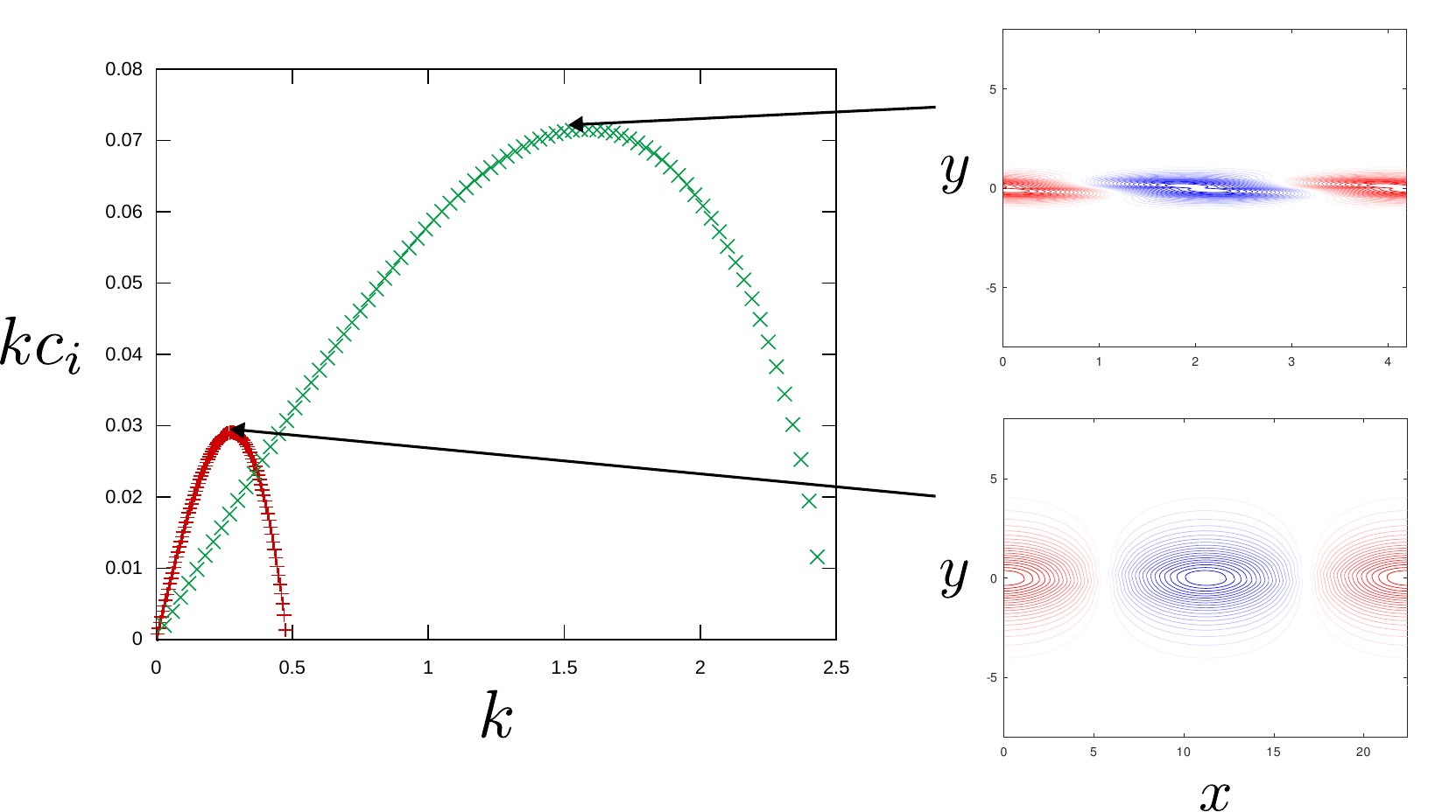}
\end{center}
\caption{
Stability analysis in the fourth quadrant of figure \ref{fig6}a, \textcolor{black}{with $U=\sin(2\pi y)$}. 
Red plus symbols: $L=16, L_d=1, a=0.5, b=2$. Green cross symbols: $L=16, L_d=1/8, a=0.5, b=1.2$.
In both cases, the most unstable eigenfunctions are indicated by arrows. \textcolor{black}{The red and blue curves are contours of the streamfunction; see figure \ref{fig7} caption.}}
\label{fourthq}
\end{figure}

\textcolor{black}{
Of particular interest from a planetary physics perspective is the fourth quadrant of figure \ref{fig6}a. 
Figure \ref{fourthq} shows the growth rates computed for the two different parameter sets, $(L,L_d,a,b)=(16,1,0.5,2)$ and $(L,L_d,a,b)=(16, 1/8, 0.5, 1.2)$. The latter setting is somewhat close to the situation found in Jupiter's atmosphere, as we shall see in Section 5.
In the chosen base flow $U=\sin(2\pi y)$, there exist many
\textcolor{black}{critical latitudes}, and in Section 6 it is shown that they must coincides with the \textcolor{black}{PV extrema} for Class (i).
For both eigenfunctions shown in figure \ref{fourthq}, one can observe that disturbances are localised near the centre of the domain, i.e., where the critical latitudes are subsonic ($M^{-1}>1$). Since $a=0.5$, other critical latitudes are supersonic ($M^{-1}<1$). 
The occurrence of vortices at subsonic latitudes is not a phenomenon specific to the current model.
Mathematically, this expectation arises from the fact that, according to Sturm's oscillation theorem, neutral modes exhibit oscillatory behaviour only when $M^{-1}$
well exceeds unity.
When $L_d$ is small, as soon as the hump of $M^{-1}$ exceeds 1, an unstable mode occurs, as noted at the end of section 3.3. This is why the eigenfunction for the $L_d=1/8$ case shown in figure \ref{fourthq} has smaller vortices.}

\subsection{Class (ii)}
\begin{figure}
\begin{center}
 \includegraphics[width=0.6\textwidth]{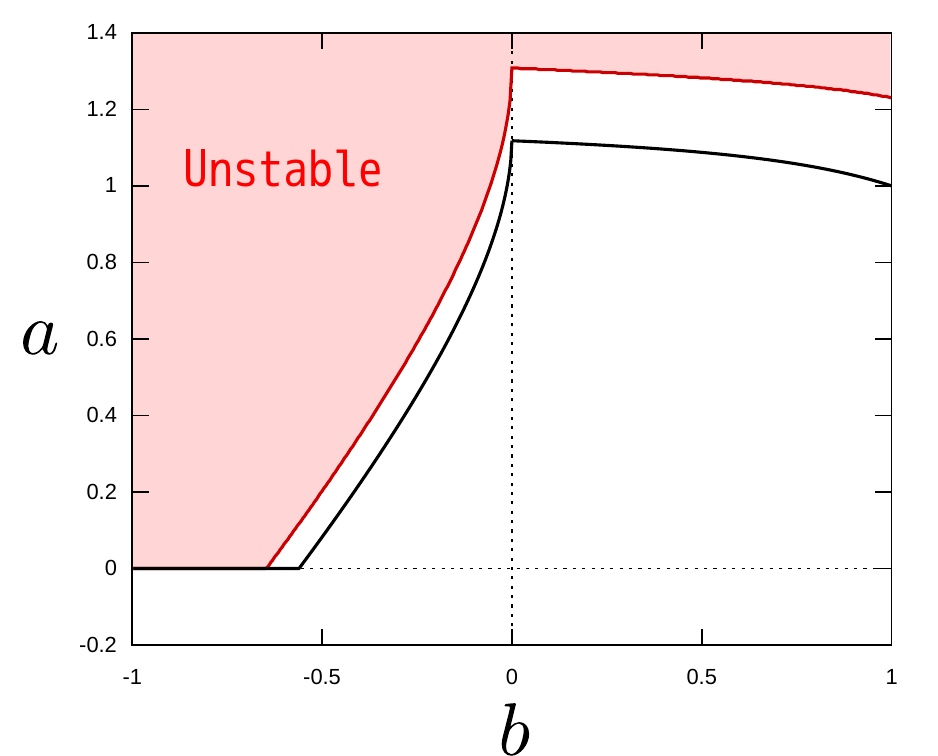}
\end{center}
\caption{
The stability of the basic flow $U=\tanh(y)$ with the parameters $L=16, L_d=1$. 
The black curve is the numerically obtained neutral curve. The red shaded area is the unstable parameter region found by Theorem \ref{ThmM}. 
}
\label{fig8}
\end{figure}

\begin{figure}
\hspace{20mm}(a)\hspace{55mm}(b)\\
\begin{center}
 \includegraphics[width=0.4\textwidth]{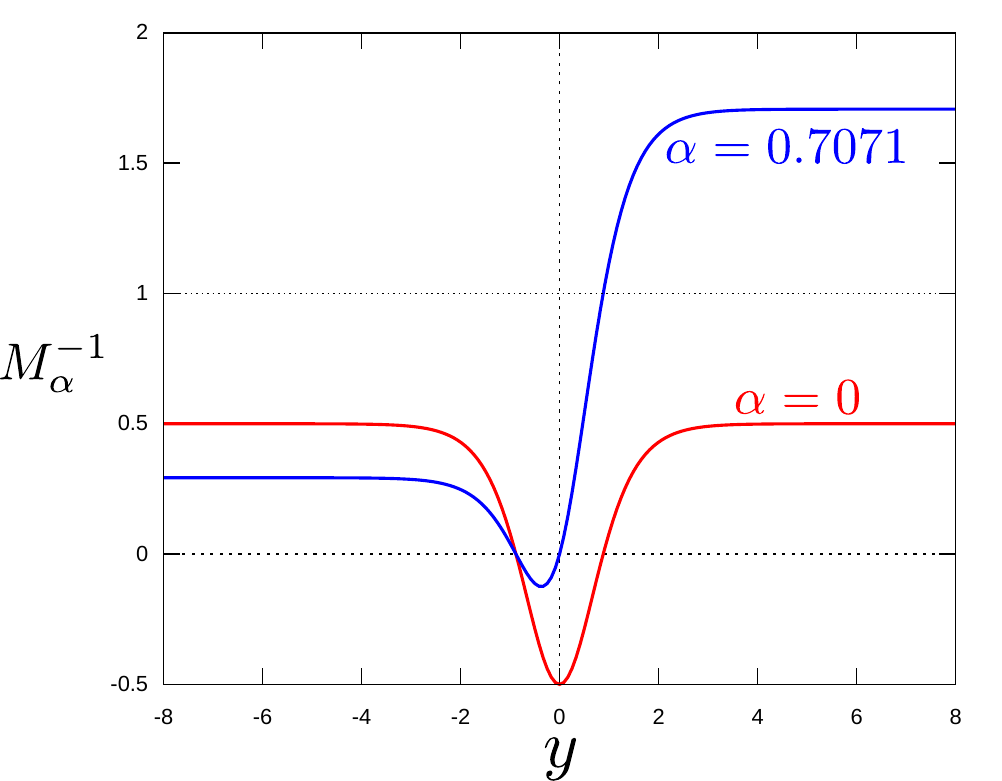}
 \includegraphics[width=0.38\textwidth]{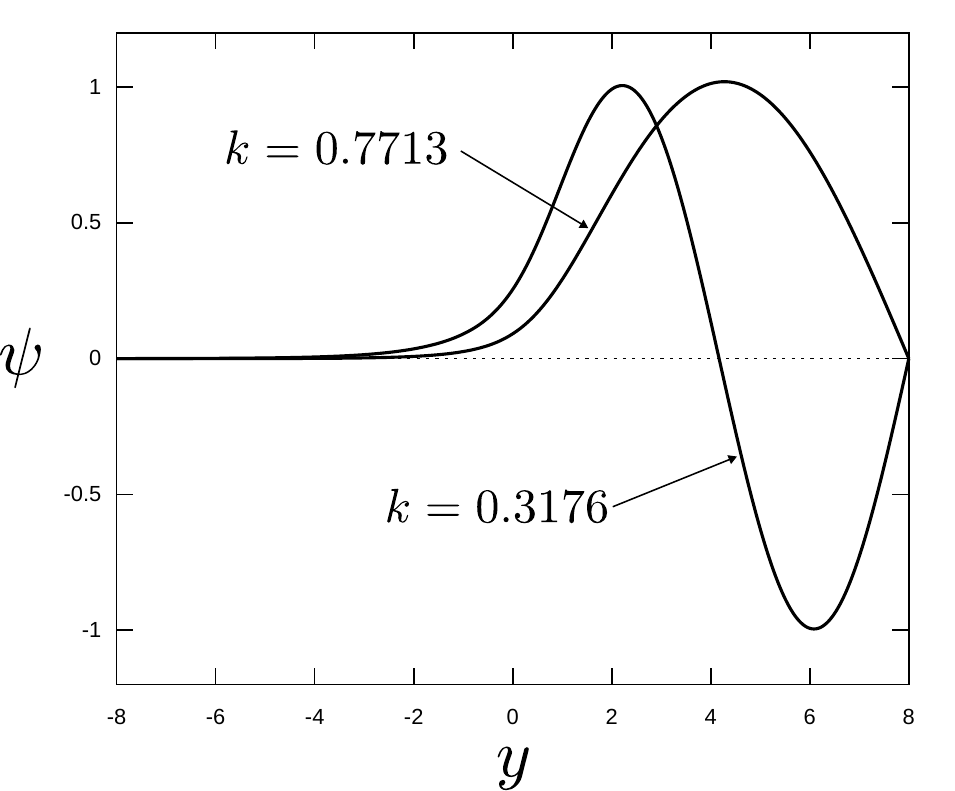}
\end{center}
\caption{(a) Reciprocal Mach number profile for $L_d=1, L=16, a=0.5,b=-0.5$. The basic flow $U=\tanh(y)$ is used. (b) The numerically obtained neutral modes at the same parameters. \textcolor{black}{They have the phase speed $c_r\approx 0.7071$.}
}
\label{fig9}
\end{figure}

\begin{figure}
\begin{center}
 \includegraphics[width=0.8\textwidth]{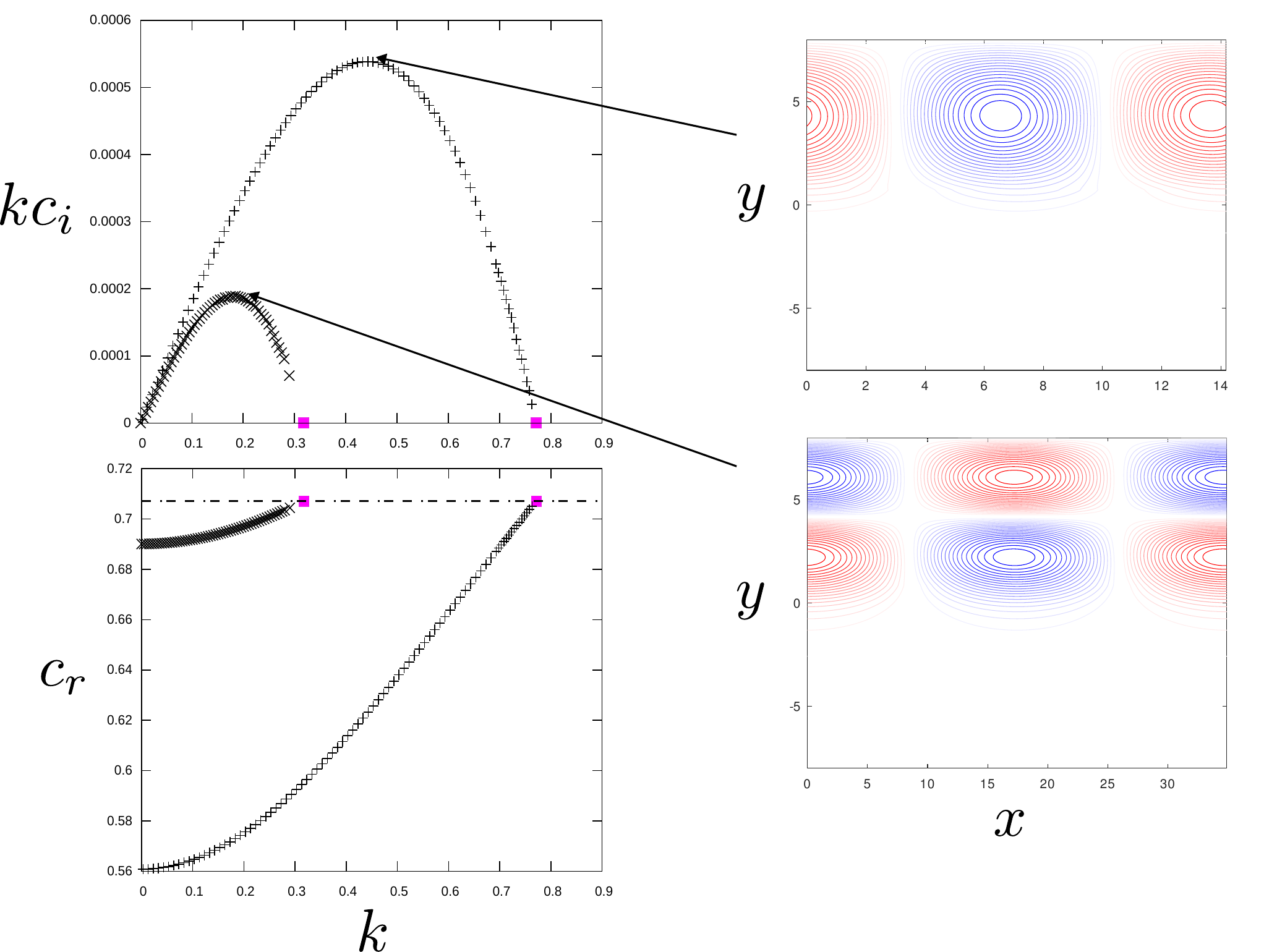}
\end{center}
\caption{The same plots as figure \ref{fig7} but for $L_d=1, L=16, a=0.5,b=-0.5$. The basic flow $U=\tanh(y)$ is used. The neutral solutions, indicated by the magenta squares, are identical to those shown in figure \ref{fig9}b. \textcolor{black}{The red and blue curves are contours of the streamfunction; see figure \ref{fig7} caption.}
}
\label{fig10}
\end{figure}

Interesting things happen when considering negative $b$ in (\ref{Qprof}). There is no longer any guarantee that the basic flow is Class (i), as the sign of the $M^{-1}_0$ profile (\ref{Mprofile}) may change. The neutral curve does depend on the choice of $U$, because the phase speed of neutral modes is not always zero. Here we mainly consider the second zonal wind profile $U(y)=\tanh(y)$, fixing $L=16, L_d=1$.  When $a$ and $b$ are positive, the neutral curve is obtained as shown in figure \ref{fig6}c. What happens if the neutral curve is extended to the region where $b$ is negative? \textcolor{black}{For example, consider the situation when $a$ is smaller than 1 and $b$ is negative; the results look like figure \ref{fig8}. 
The emergence of the unstable region is due to the modes with non-zero phase speed, because clearly $M_0^{-1}$ will be everywhere smaller than 1.}


It is easy to see that the \textcolor{black}{PV extrema} are zeros of $U(y)$ and $\{a+(b-a)\, \text{sech}^2 (y)\}$. For example, consider the case $a=0.5, b=-0.5$. The zeros of $\{a+(b-a)\, \text{sech}^2 (y)\}$ are at $\pm 0.8814$, and thus the \textcolor{black}{PV extrema} are $Y_Q=\{-0.8814,0,0.8814\}$. One of these three points must coincide with the zero of the $U(y)-\alpha$, for $M^{-1}_{\alpha}$ to be continuous. Thus there are neutral modes with $c_r=0$ and $c_r\approx \pm U(0.8814)\approx \pm 0.7071$. 
If $\alpha=0$ is chosen, the $M^{-1}_{\alpha}$ profile is everywhere less than unity as shown by the red curve in figure \ref{fig9}a. Hence, the flow is stable for steady modes. However, when $\alpha=0.7071$ is used, $M^{-1}_{\alpha}$ exceeds 1 significantly, as indicated by the blue curve in \ref{fig9}b. 

Travelling wave type neutral modes with a phase speed of 0.7071 indeed appear for the parameter choice $a=0.5, b=-0.5$ (figure \ref{fig9}b). 
For $\alpha=0.7071$, Class (ii) conditions are satisfied. This follows from the monotonicity of $U(y)$; see the comments below (\ref{union}).
As would be expected from these facts, unstable modes appear when $k$ is reduced from the neutral values (figure \ref{fig10}). The eigenfunctions with the largest growth rates are similar to the neutral mode, with vortices concentrated in regions where $M^{-1}_{\alpha}$ exceeds 1.

\textcolor{black}{The fact that the neutral curve for $b<0$ depends on $U$ suggests that the situation is much more complicated when the basic flow is not of Class (i). For example, if the basic flow $U(y)=\sin(2\pi y)$ is used,} the only reference shift $\alpha$ that would make $M^{-1}_{\alpha}$ continuous is 0, when $a>0, b<0$, and $L$ is large. However, this does not mean that considering only steady modes is sufficient. 
The reason for this is that when $\{U(y)|y\in \Omega_- \}\cap\{U(y)|y\in \Omega_+ \}$ is non-empty, there may be a singular neutral mode (see the remark below (\ref{phasei})).
Such singular modes are beyond the scope of this paper, \textcolor{black}{and in fact there is no need to consider them in the planetary context as long as we assume that Jupiter's atmosphere is of Class (i).}

\section{Implications of the model results to planetary atmospheres}

\textcolor{black}{In this section, we summarise implications of our model analysis in geophysics problems, including deep jets on Jupiter and Saturn and the inference of Saturn’s rotation period (which is otherwise elusive because its magnetic field is not tilted). We then motivate nondimensional, idealised cases and show the deep jets associated with neutral stability are reasonable, even with slightly varying reciprocal Rossby Mach number. We end the section with a few comments about weakly unstable scenarios for the neighbourhoods of prominent features such as Jupiter’s Great Red Spot and Saturn’s Polar Hexagon and Ribbon. }


Before delving into the analysis of the model introduced in the last section, we emphasise that our purpose here is not to faithfully reproduce the details of Jupiter's atmospheric phenomena, but rather to elucidate \textcolor{black}{key physical mechanisms.
It is understood that the beta-plane \textcolor{black}{$1\frac{3}{4}$} quasi-geostrophic model is derived from a series of simplifications, for example, the actual atmospheres of Jupiter and Saturn have continuous vertical-structure.
Also, the value of $L_d$} may vary with latitude and we do not have good estimates in hand outside the vicinity of the latitude range shown in figure 3. 
\textcolor{black}{The assumption that Jupiter's atmosphere falls under Class (i) is inferred from observations, with the understanding that the actual planet is dynamically active, with long periods of stability punctuated by episodic storm outbreaks.} 
Remote-sensing observations are accompanied by noise, and when substituting the figure 1 data into (3.3), it is evident that, regardless of how we choose the zonal wind shift, $M^{-1}_{\alpha}(y)$ cannot be made into a continuous function. 


Nevertheless, there is a consensus among multiple independent research groups that the observed $Q’(y)$ and $U(y)$ tend to have the same sign on Jupiter and Saturn \textcolor{black}{\citep{Dowling1993,Read2006,Read2009,Marcus2011}}, and when this property is ideal, the basic flow falls under Class (i).
\textcolor{black}{
As a minimum check to test that the link between the stability theory and the correlation seen in figure 3 is robust with respect to noise, the following numerical experiment was performed. Given the base flows, regardless of their class, the value of $L_d$ that \textcolor{black}{makes the flow configuration neutral}
can be computed by the eigenvalue problem (\ref{EQ}). Therefore, if $\kappa_0$ is computed from that $L_d$, we can compare $\kappa_0^{-2} Q'$ and $U$ to check their correlation. 
We spline interpolated $U$ and $Q_y$ in the latitude range shown in figure 3, corresponding to $L \approx 10860 \, \rm km$. 
The numerical eigenvalue problem (\ref{EQ}) yields $\kappa_0^{-1}\approx 1700 \, \rm km$, which is reasonably close to the value $\kappa_0^{-1}= 1750 \, \rm km$ used in figure 3. Moreover, the neutral wave has a relatively small phase speed 
$c_r\approx 6 \, \rm m s^{-1}$. As a consequence, plotting $(U - c_r)$ and $\kappa_0^{-2}Q'$ with the computed $\kappa_0^{-1}, c_r$ gives the same level of correlation as in figure 3. This result is robust with respect to the artificially set boundary conditions in the computation.
}



\begin{figure}
\begin{center}
  \includegraphics[width=0.6\textwidth]{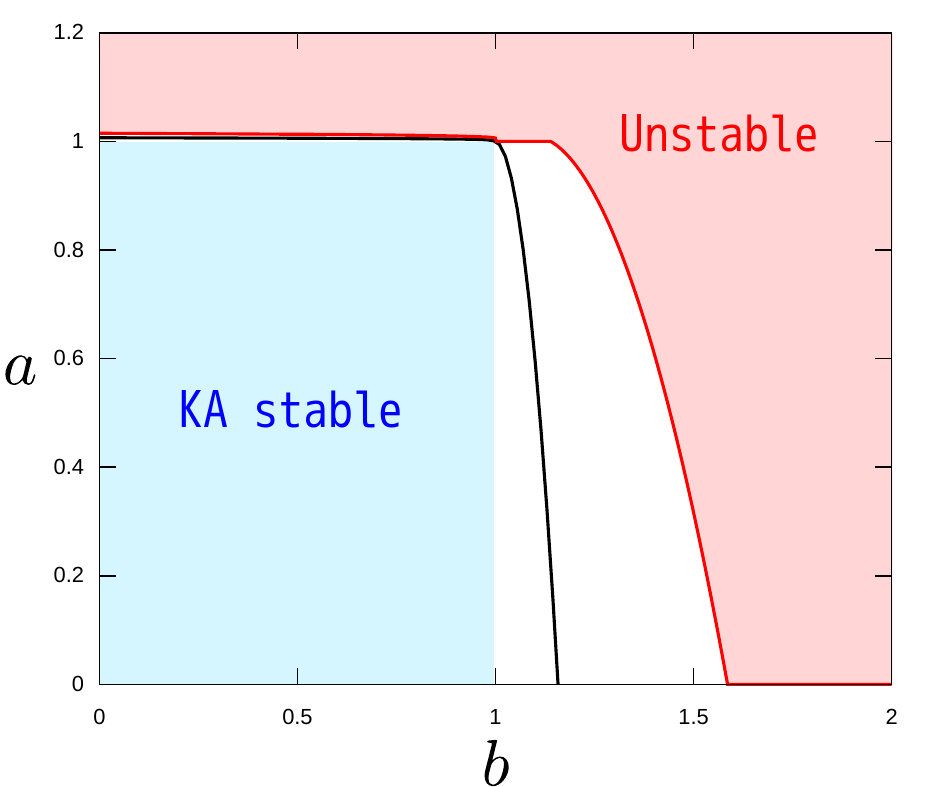}
\end{center}
\caption{
The black curve is the neutral curve for $L_d =1/2\pi$, $L=16$. The basic flow profiles (\ref{Qprof}) and (\ref{C1B1}) are used. The blue shaded region is stable from Theorem 1. The red shaded region is unstable from Theorem 2.
}
\label{fignew1}
\end{figure}

On Jupiter, the jet wavelength is $L_U \approx 2\pi L_d$, as can be seen from figure \ref{fig:Read2006_profiles} using the fact that $1^\circ$ latitude is approximately $1200 \, \rm km$; a similar relationship holds for Saturn. Based on this observation, we choose $L_d =1/2\pi$ in the nondimensional model, employing the profiles (\ref{Qprof}) and (\ref{C1B1}). 
The black curve in figure \ref{fignew1} represents the numerically obtained neutral curve; as expected, it lies between the stable and unstable regions obtained by Theorems 1 and 2. The hurdle theorem result now better approximates the neutral curve than figure 7c. 
Also, figure \ref{fignew1} clearly demonstrates that even a slight deviation from the KA-II stability boundary may result in flow instability. 
As discussed in Section 1, the introduction of the Rossby Mach number was motivated by the physical interpretation of KA-II. Our discovery that KA-II is relatively sharp under Jupiter's atmospheric conditions reinforces the validity of that physical mechanism.


Stabilising alternating jets with critical latitudes in a gas-giant weather layer, or in an analogous \textcolor{black}{$1\frac{3}{4}$} layer system, via KA-II requires that there be alternating jets in the deep layer. Such deep jets must in turn be stabilised by some physical process other than KA-II. The key there appears to be that the deep jets operate in a quite different geometry---that of a rapidly rotating deep sphere instead of a shallow spherical shell, as investigated by \cite{Ingersoll_Pollard1982}. Regardless of the physics behind the synchronisation between the weather and deep jets, it is a fortunate circumstance for geophysicists because in practice, to detect stable critical latitudes in a weather layer is to infer deep jets, which is how Jupiter’s deep jets were first discovered, decades before the \textit{Juno} gravity mission confirmed their existence \citep{Dowling1989,Dowling1993,Dowling1995_estimate}. Our take is that KA-II provides the weather layer jets with the flexibility to adjust to alternating deep jets, and 
\textcolor{black}{this represents a meaningful step forward in understanding the overall stability of the atmosphere-interior system.}

\begin{figure}
\hspace{15mm}(a)\hspace{53mm}(b)\\
\begin{center}
  \includegraphics[width=0.9\textwidth]{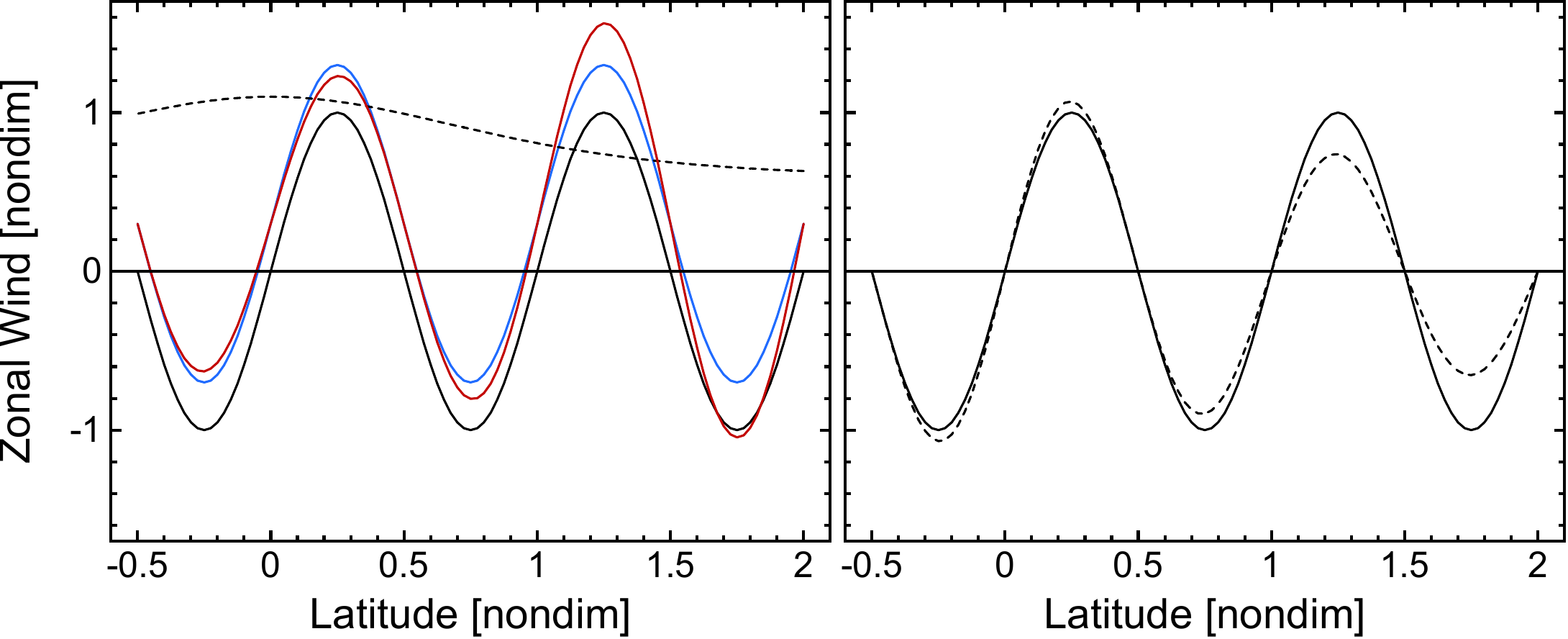}
\end{center}
\caption{
\textcolor{black}{Idealised, nondimensional zonal wind profiles relevant to Jupiter. 
Note that the domain of latitudes shown is a zoom-in near the origin of the full domain, and $L/L_d \gg 1$ is assumed such that $\kappa_0^{-2}\approx L_d^2$. 
(a) The solid black curve is $U(y) = \sin(2\pi y)$. The blue curve is $U_{\rm deep}(y)$ corresponding $M^{-1} = 1$, which amounts to a positive shift by $\beta = 0.3$. The dashed black curve is the profile $M^{-1}(y) = a+(b-a){\rm sech}^2(y)$, with $(a,b) = (0.6,1.1)$, which is a point on the neutral curve shown in figure \ref{fignew1}; the red curve is the corresponding $U_{\rm deep}(y)$.  (b) Profiles of $U$ (solid) and $\kappa_0^{-2}Q_y $ (dashed) corresponding to the dashed $M^{-1}$ and red $U_{\rm deep}$ profiles in (a); compare with figure \ref{fig:Read2006_profiles2}. }
}
\label{fignew2}
\end{figure}

Assuming that Jupiter's atmosphere favours neutral states, the deep layer profile $U_{\rm deep}(y)$ can be calculated from (2.4). 
In the previous studies, zonal-wind pairs (weather layer and deep layer) appropriate for the \textcolor{black}{$1\frac{3}{4}$} layer model applied to Jupiter were calculated from Voyager winds and vorticity data and plotted in \cite{Dowling1989, Dowling1995_estimate, Dowling2020}.  The deep-layer westward jets tend to be similar to the cloud-top westward jets, whereas the deep-layer eastward jets tend to be stronger than the weather-layer eastward jets by about 50\% \citep{Dowling1995_estimate}. 
For our model where Jupiter’s weather-layer profile is idealised as a sinusoid, $U(y) = \sin(2\pi y)$, if we take $M^{-1} = 1$ (i.e. $a=b=1$) with $\kappa_0^{-1} \approx L_d$, (2.4) yields the dimensional deep jet profile $U_{\rm deep}(y) = U(y)+\beta L_d^2/U_0$, such that the deep-layer profile is the same sinusoid but with a positive shift. The size of this shift can be estimated for \textcolor{black}{Jupiter assuming $\beta \approx 3.5\times10^{-12} \, \rm m^{-1} s^{-1}$, $L_d \approx 1750 \, \rm km$ and a stratospheric wind amplitude, $U_0 \approx 25 \, \rm m \, s^{-1}$, as in figure \ref{fig:Read2006_profiles2}, which yields $\beta L_d^2/U_0 \approx 10.7/25 \approx 0.4$. Alternatively, the \textcolor{black}{$1\frac{3}{4}$} layer model applied to Jupiter's Great Red Spot, which is centred closer to the equator at latitude $-23^\circ$, would typically use $L_d \approx 2000 \, \rm km$ and a tropospheric wind amplitude, $U_0 \approx 50 \, \rm m \, s^{-1}$, which yields $\beta L_d^2/U_0 \approx 14/50 \approx 0.3$; both are consistent with  Voyager results \citep{Dowling1995}}. The non-dimensionalised $U_{\rm deep}(y)$ \textcolor{black}{with $\beta L_d^2/U_0 = 0.3$} is depicted in figure \ref{fignew2}a by the blue curve. We can also calculate $M^{-1}$ using other points from the neutral curve, for example, the neutral point $(a,b) = (0.6,1.1)$ yields the $U_{\rm deep}(y)$ profile represented by the red curve in figure \ref{fignew2}a.  The family of neutral solutions provides helpful information about expected variations in observations.

Recall that for Class (i), the choice of $\alpha$ is unique. Furthermore, on the neutral curve, the value of $\alpha$ must equal to the phase speed of the neutral modes (to be shown in Theorem 4), consistent to the observation by \cite{Read2009b} that long-wavelength Rossby waves in Saturn have the same $10^{\rm h} 34^{\rm m}$ rotation period. 
Consequently, for those neutral modes, the \textcolor{black}{PV extrema and critical latitudes must coincide.} As seen in the previous section, when $L_d$ is small, the middle critical latitude in the model is nearly sonic ($M^{-1}=1$), when the flow is nearly neutral. This situation is consistent with the observations by \cite{Read2006,Read2009} where it is revealed that Jupiter and Saturn each have at least a dozen stable or marginally stable (i.e., supersonic or sonic, $M^{-1} \leqslant 1$) critical latitudes, consistent with the fact that the alternating jets on those planets persist on a decadal timescale \citep{Porco2003}.

%

Also, as first spotted by \cite{Dowling1989} and later confirmed by \cite{Marcus2011}, Jupiter's Great Red Spot straddles a critical latitude. Numerical experiments \citep[e.g.,][]{Dowling1993} suggest that vortices appear at subsonic ($M<1$) latitudes, as theoretically expected for nearly neutral unstable modes. As seen in figure \ref{fig:Read2006_profiles}, the nature of the correlation of $U$ and $Q_y$ differs between the latitude ranges of $-30^\circ$ to $-20^\circ$ and $-50^\circ$ to $-30^\circ$, and the former range is where features like the Great Red Spot and Oval BA are observed. As a useful exercise, one can plot $M^{-1}(y)$ profile using the data shown in figure 1 and $L_d$ estimated in figure 3, to demonstrate the presence of a hurdle that aligns with the range of latitudes $-30$ to $-20$ degrees, and that at other latitudes, $M^{-1}(y)$ is characterised by several humps that peak close to unity.  


Further to this point, \cite{Dowling2020} pointed out that Saturn has two different, persistently wavy jets sandwiched between straight jets in its northern hemisphere, the Ribbon and the Polar Hexagon, and offered the hypothesis that these are examples of ‘subsonic’ regions sandwiched between ‘sonic’ or ‘supersonic’ regions. Although the \textcolor{black}{temperature} fields necessary to empirically determine the stretching vorticity, and hence the PV field, at and below Saturn's cloud tops are difficult to obtain with the data in hand, the morphology of these wavy jets is well established, which suggests that an analysis couched in terms of $M^{-1}(y)$ along the lines outlined here would be instructive.


\section{Mathematical proofs}

Our strategy to show sufficient conditions for instability (i.e. necessary conditions for stability) is similar to the two-step procedure in \cite{Howard1964} and \cite{Lin2003}: first prove the conditions that guarantee the existence of a neutral solution, and then establish the existence of unstable modes in its neighbourhood in parameter space. A regular mode, to be defined below, must exist for these processes to take place, and guaranteeing this essentially corresponds to the second Class (ii) condition.

\subsection{Choice of $\alpha$ for Class (i) basic flows}

We first note that the following theorem holds.
\begin{thm}\label{classi}
Suppose the basic flow is Class (i). Then there is only one $\alpha \in \mathcal{R}$ that makes $M_{\alpha}^{-1}$ continuous. Moreover, $M_{\alpha}^{-1}\neq 0$ for all $y\in \Omega$.
\end{thm}


This theorem can be shown as follows. 
\textcolor{black}{The possibility of $M^{-1}_{\alpha}$ vanishing only occurs at $y_l \in Y_Q$. 
For Class (i), $U'(y_l)\neq 0$ for all $y_l \in Y_Q$, because otherwise $M_{\alpha}^{-1}$ becomes singular, in view of the assumption that $Q''(y_l)\neq 0$ for all $y_l \in Y_Q$. 
L'Hôpital's rule now suggests that $M_{\alpha}^{-1}(y_l)=Q''(y_l)/U'(y_l) \neq 0$. 
We can also show that the $\alpha \in \RealN$ that makes $M^{-1}_{\alpha}$ continuous and one-signed is uniquely determined. 
Suppose there are two possible values of $\alpha$, $\alpha_1$ and $\alpha_2$, say. Since $M_{\alpha_1}^{-1}$ and $M_{\alpha_2}^{-1}$ are continuous functions that do not vanish in $\Omega$, $(M_{\alpha_1}-M_{\alpha_2})$ is a continuous function. However, this implies that $\frac{\alpha_2-\alpha_1}{Q'}$ is a continuous function, which is not possible unless $\alpha_1=\alpha_2$.
}

Class (i) is an extension of class $\mathcal{H}$ in \cite{Howard1964}, and class $\mathcal{K}^+$ in \cite{Lin2003} to the quasi-geostrophic equation. However, such strong restrictions for basic flows are not necessary to derive the hurdle theorem, as remarked in Section 3.

\subsection{Classification of neutral modes}

Let us consider the neutral solutions, setting $c_i=0$. The key to classify neutral modes is to note that they can potentially become singular at the \textcolor{black}{critical latitudes,} the points at which $U(y)$ coincides with the phase speed $c_r$. Mathematically, the set of the \textcolor{black}{critical latitudes}, $Y_{U,c_r}=\{y \in \Omega | U(y)=c_r\}$, are regular singular points of (\ref{EQ}) when $c_i=0$. The appropriate tool for analysing the behaviour of solutions around such singularities is Frobenius' method, from which it can be shown that $\psi$ is continuous but $\psi'$ may be discontinuous at the \textcolor{black}{critical latitudes} \citep{Lin1955}. A necessary and sufficient condition for no jumps to exist at a \textcolor{black}{critical latitude} is that \textcolor{black}{either $Q'=0$ or $\psi=0$, or both,} occur there. Here, following \cite{Drazin_Reid1981}, we briefly explain the nature of \textcolor{black}{the} \textcolor{black}{critical latitudes} without going into the details of Frobenius' method.

Because of the singularities, neutral solutions must be understood as the vanishing $c_i$ limit in unstable solutions. Let us multiply (\ref{EQ}) by $\psi^*$ and take the imaginary part, keeping finite $c_i$:
\begin{eqnarray}
\psi^*\psi''-\psi \psi^*\, ''
+\frac{2ic_i Q'}{(U-c_r)^2+c_i^2}|\psi|^2=0 \, .
\label{RS}
\end{eqnarray}
Integrate (\ref{RS}) across a \textcolor{black}{critical latitude}, $y=y_c$ (i.e., $U(y_c)=c_r$), and take the limit $c_i\rightarrow 0+$,
\begin{eqnarray}
[\psi^*\psi'-\psi \psi^*\, ']^{y_c+}_{y_c-}&=&\lim_{\epsilon\rightarrow 0+}\lim_{c_i\rightarrow 0+}\int^{y_c+\epsilon}_{y_c-\epsilon} \frac{-2ic_i Q'}{(U-c_r)^2+c_i^2}|\psi|^2 dy
\label{jump}
\end{eqnarray}
An intuitive way to \textcolor{black}{evaluate the right hand side} is as follows. 
If $\epsilon>0$ is sufficiently small, we may be able to assume that $Q'|\psi|^2$ is almost a constant, and that $U-c_r$ is approximately $U_c'(y-y_c)$, where $U'_c=U'(y_c)$. Then the right hand side of (\ref{jump}) can be explicitly worked out by using the well-known formula that can be found by differentiating the arctangent function:
\begin{eqnarray}
-2i(Q'|\psi|^2)|_{y=y_c}\lim_{\epsilon\rightarrow 0+}\lim_{c_i\rightarrow 0+}\frac{2}{U_c'}\arctan\left (\frac{\epsilon U_c'}{c_i}\right )=-2i\pi\left. \left (\frac{Q'|\psi|^2}{|U'|}\right )\right |_{y=y_c}.
\end{eqnarray}
\textcolor{black}{The above argument is consistent with the viscous problem when the singularity of the inviscid solution is regularised by viscosity \citep{Lin1955}. 
The regularisation by inertia is also possible (\cite{Haberman1972,Robinson1974}) -- although such a situation is relevant to nonlinear equilibrium states (e.g. \cite{DeguchiWalton2018}), here we only consider the linear problem.}

The above result suggests that if multiple \textcolor{black}{critical latitudes} appear as $Y_{U,c_r}=\{y_1, y_2, \dots, y_N\}$, integrating (\ref{RS}) over $\Omega$ yields
\begin{eqnarray}
0=\sum_{j=1}^{N}\left. \left (\frac{Q'|\psi|^2}{|U'|}\right )\right |_{y=y_j}.
\label{phasei}
\end{eqnarray}
If $Q'=0$ or $\psi=0$ happen at all the \textcolor{black}{critical latitudes}, there are no jumps at all, and hence the neutral solution $\psi$ is real and $C^2_0$. Here, it is convenient to define terminology to distinguish between neutral-mode types: 
\smallskip
\begin{description}
   \item[Pathological mode]~ empty $Y_{U,c_r}$ or $\psi$ vanishes at all \textcolor{black}{critical latitudes};

   \item[Regular mode]~ eigenfunction is not pathological and is $C^2_0$;

   \item[Singular mode]~ eigenfunction is not $C^2_0$.
\end{description}
\smallskip
\textcolor{black}{In standard textbooks like \cite{Drazin_Reid1981} and in previous research, there has been no distinction made between pathological modes and regular modes. We shall show in Section 6.3 that when the first Class (ii) condition is satisfied and the reciprocal Rossby Mach number surpasses a hurdle, neutral solutions exist with some choices of $k$, and at least one of them must be a regular mode. Moreover, if the second Class (ii) condition holds, unstable modes must exist around the (non-pathological, least oscillatory) regular neutral mode (Section 6.4). }

If the intersection of sets $\{U(y)|y\in \Omega_+ \}$ and $\{U(y)|y\in \Omega_- \}$ is non-empty, the summation in (\ref{phasei}) may cancel, and hence a singular mode may exist (thus for Class (i) there are no singular modes). For the singular modes we cannot use the standard Sturm-Liouville theory to be used in sections 6.3 and 6.4, and for the pathological modes we have difficulty in showing neighbourhood instability.



\subsection{A sufficient condition for existence of a neutral mode}

Let us assume that for some $\alpha\in \RealN$ the reciprocal Rossby Mach number $M^{-1}_{\alpha}(y)$ becomes continuous. 
Write $\lambda=-k^2$.
Then from (\ref{EQ}) the neutral modes with the phase speed $c=\alpha$ must satisfy
\begin{eqnarray}
-\psi'' + (L_d^{-2}-\kappa_{0}^2M_{\alpha}^{-1})\psi=\lambda \psi, \qquad y \in (-L/2,L/2),\label{EQ2}
\end{eqnarray}
and the Dirichlet boundary conditions.
This is a regular Sturm-Liouville problem with eigenvalue $\lambda$. It is well-known that all eigenvalues are real, and they can be ordered as $\lambda_0<\lambda_1<\lambda_2<\dots$, where $\lambda_n \rightarrow \infty$ as $n\rightarrow \infty$. By integration by parts, it is easy to check that the associated eigenfunctions, $\psi_0, \psi_1,\psi_2,\dots$, are real and satisfy the equation
\begin{eqnarray}
R_{\alpha}(\psi_n)\equiv \frac{\int_{\Omega} \{|\psi_n'|^2+(L_d^{-2}-\kappa_{0}^2M_{\alpha}^{-1})|\psi_n|^2\}dy}{\int_{\Omega} |\psi_n|^2dy}=\lambda_n.\label{Rquo}
\end{eqnarray}
Here  the Rayleigh quotient, $R_{\alpha}$, depends on $\alpha$. It should also be noted from Sturm's oscillation theorem that the $n$th eigenfunction $\psi_n$ has $n$ zeros in the interval $(-L/2,L/2)$, a useful fact to be used below.
\textcolor{black}{In figure 6a, the red curve is $\psi_0$, and the other curves may be $\psi_1, \psi_2, \psi_3$. Likewise the modes shown in figure 11b are the zeroth and first modes. }

The minimum eigenvalue $\lambda_0$ can be found by the optimisation problem
\begin{eqnarray}
\lambda_0=\min_{\phi\in C^2_0}R_{\alpha}(\phi),\label{mini}
\end{eqnarray}
where the unique minimiser is the zeroth eigenfunction $\phi=\psi_0$. Note that there is no problem to extend the search space to
\begin{eqnarray}
H_0^1=\left \{\phi \left | \int_{\Omega}|\phi|^2dy<\infty, \int_{\Omega}|\phi'|^2dy<\infty, \phi(-L/2)=\phi(L/2)=0\right. \right \}
\end{eqnarray}
by a density argument.
Here, following the usual notation in mathematics, $H$ implies that the space is a Hilbert space, the superscript $1$ implies that the square integrability of the first derivative, and the subscript $0$ implies that the Dirichlet boundary conditions are satisfied.

A sufficient condition for existence of a neutral mode is then summarised as follows.
\begin{thm}\label{neutralG}
Suppose there is $\alpha \in \RealN$ that makes $M^{-1}_{\alpha}$ continuous.
If there exists $g(y)\in H^1_0$ such that $R_{\alpha}(g)\leq 0$, there is a neutral mode with the phase speed $c_r=\alpha$.
\end{thm}
If the assumption of the theorem is met, a neutral mode can be found because $\lambda_0=\min_{\phi\in H_0^1}R(\phi) \leq R(g) \leq 0$ implies that the minimiser of (\ref{mini}) is the neutral eigenfunction having the wavenumber $k=k_0\equiv \sqrt{-\lambda_0}\geq 0$. \textcolor{black}{This neutral mode is $\psi_0$ introduced earlier. The neutral curves shown in figures 7a, 10, 13 are indeed determined by $\psi_0$, and crucially, this mode must be a regular mode.}

\textcolor{black}{
Since $c_r=\alpha$ for $\psi_0$, the \textcolor{black}{critical latitudes} (the points $y_j$ at which $U(y_j)=c_r$ is satisfied) must appear on the \textcolor{black}{PV extrema} (i.e. $Y_{U,c_r} \subset Y_Q$). 
However, for Class (i) basic flows, the stronger result $Y_Q=Y_{U,c_r}$ can be shown for $\psi_0$, because $Y_Q=Y_{U,\alpha}$ as remarked just below (\ref{mathscrR}), and there is only one choice of $\alpha$ from Theorem 3. 
}

 

In the next section, we shall show the following theorem:
\begin{thm}\label{unstable}
Suppose the basic flow is Class (ii), and fix $\alpha$ so that the Class (ii) conditions are satisfied. If there exists $g(y)\in H^1_0$ such that $R_{\alpha}(g)< 0$, the flow is unstable. 
Furthermore, if the basic flow is Class (i), a necessary and sufficient condition for stability is that no such $g(y)\in H^1_0$ exists.
\end{thm}
\noindent
The second half of the theorem is somewhat similar to that derived by \cite{Howard1964} and \cite{Lin2003} for shear flows, but the first half is entirely new. We also remark that in the case of the Rayleigh equation, it can be proved that the phase speed of the neutral solution is in the range of $U(y)$, denoted $\mathcal{R}$, by Howard's semicircle theorem, and that there are no pathological modes when $k>0$. 
\textcolor{black}{Moreover, when $k=0$, a regular mode can be found analytically.}
Those facts are used, for example, in proofs by \cite{Howard1964}, \cite{Balmforth_Morrison1999}, and \cite{Hirota_etal2014}, but they do not in general apply to stability problems more complex than the Rayleigh equation. This is essentially the reason why \cite{Tung1981} struggled to incorporate the effect of non-zero $\beta$ into the theory, but as we will see in the next section, the solution is, in fact, simple.

\subsection{Existence of an unstable mode}
\label{appendix:existence_unstable_mode}

Here we show Theorem \ref{unstable}. The proof is rather technical and readers who are not interested in the mathematical details may skip this subsection without losing the thread of the discussion. We first note that upon writing $\gamma\equiv -L_d^{-2}-k^2=\lambda-L_d^{-2}$, (\ref{EQ}) becomes
\begin{eqnarray}
\psi''+\left (\frac{Q'}{U-c}+\gamma \right ) \psi =0.\label{psieq0}
\end{eqnarray}
The corresponding dispersion relation, $F(c,\gamma)=0$, can be formulated by two linearly independent solutions of (\ref{psieq0}). They are analytic functions in the upper or lower half complex $c$-plane, and behave regularly with respect to $\gamma$, and so does $F(c,\gamma)$. 
A neutral solution is obtained when $c_i$ is brought close to zero in the dispersion relation.
From the symmetry of the inviscid problem, neutral solutions always exist as pairs, i.e. 
the $c_i=0+$ mode and the $c_i=0-$ mode. 

Let us suppose that the assumptions of Theorem \ref{unstable} are satisfied. From Theorem \ref{neutralG} we know that there is a \textcolor{black}{regular} neutral mode $\psi_0$ with wavenumber $k_0=\sqrt{-\lambda_0}>0$ and phase speed $\alpha$. This mode satisfies
\begin{eqnarray}
\mathcal{L}\psi_0\equiv \psi_0''+\left (\frac{Q'}{U-\alpha}+\gamma_0 \right ) \psi_0=0.
\label{L0}
\end{eqnarray}
Here, we define the linear operator $\mathcal{L}$ for later use.

In view of the argument \textcolor{black}{just below (\ref{psieq0})}, we can compute the coefficients of the Taylor expansion of $c(\gamma)$ around the neutral mode in the upper half complex plane.
Writing $\gamma_0\equiv \lambda_0-L_d^{-2}$, around the neutral mode, the following expansion holds,
\begin{eqnarray}
c_i(\gamma)=\left. \frac{dc_i}{d\lambda}\right |_0(\gamma-\gamma_0)+O((\gamma-\gamma_0)^2) \, .
\label{Tay1}
\end{eqnarray}
Here and hereafter, the symbol `$|_0$' attached to a quantity means that it is evaluated at the $c_i=0+$ neutral mode. The expression (\ref{Tay1}) is valid as long as $c_i$ does not become negative. Our goal below is to show $\left. \frac{dc_i}{d\gamma}\right |_0\neq 0$; then (\ref{Tay1}) implies that an unstable mode can be found by slightly varying $\gamma$ from the neutral value.  

Differentiate (\ref{psieq0}) \textcolor{black}{with respect to} $\gamma$,
\begin{eqnarray}
\psi_{\gamma}''+\left (\frac{Q'}{U-c}+\gamma \right ) \psi_{\gamma} 
+\left (1+\left (\frac{Q'}{U-c}\right )_{\gamma} \right ) \psi=0 \, ,
\label{diff}
\end{eqnarray}
where the subscript $\gamma$ denotes the differentiation.
Evaluate this equation at the neutral parameter,
\begin{eqnarray}
\mathcal{L} \psi_{\gamma}|_0 
+\left (1+\left . \left (\frac{Q'}{U-c}\right )_{\gamma}\right |_0 \right ) \psi_0=0 \, .
\label{L1}
\end{eqnarray}
Now, combine (\ref{L0}) and (\ref{L1}) to obtain
\begin{eqnarray}
0=\int_{\Omega}(\psi_0\mathcal{L} \psi_{\gamma}|_0-\psi_{\gamma}|_0\mathcal{L} \psi_0)dy
+\int_{\Omega}\left (1+\left . \left (\frac{Q'}{U-c}\right )_{\gamma}\right |_0 \right ) \psi_0^2dy \, .
\label{intLL}
\end{eqnarray}
The first integral vanishes after integration by parts. Therefore
\begin{eqnarray}
\left. \frac{dc}{d\gamma}\right |_0=-\frac{\int_{\Omega} \psi_0^2dy }{K}=-\frac{(K_r-iK_i)\int_{\Omega} \psi_0^2dy }{K_r^2+K_i^2} \, ,
\label{dcdg}
\end{eqnarray}
where
\begin{eqnarray}
K=K_r+iK_i=\lim_{c_i \rightarrow 0+}\int_{\Omega} \frac{Q'\psi_0^2}{(U-\alpha-ic_i)^2}dy \, .
\label{Klim}
\end{eqnarray}
The limit can be worked out as
\def\Xint#1{\mathchoice
   {\XXint\displaystyle\textstyle{#1}}%
   {\XXint\textstyle\scriptstyle{#1}}%
   {\XXint\scriptstyle\scriptscriptstyle{#1}}%
   {\XXint\scriptscriptstyle\scriptscriptstyle{#1}}%
   \!\int}
\def\XXint#1#2#3{{\setbox0=\hbox{$#1{#2#3}{\int}$}
     \vcenter{\hbox{$#2#3$}}\kern-.5\wd0}}
\def\ddashint{\Xint=}
\def\dashint{\Xint-}
\begin{eqnarray}
K=\dashint_{\Omega} \frac{Q'\psi_0^2}{(U-\alpha)^2}dy+i\pi\kappa_0^2\sum_{j=1}^N \left. \left (\frac{M_{\alpha}^{-1}\psi_0^2}{|U'|} \right) \right|_{y=y_j} \, ,
\label{Kri}
\end{eqnarray}
noting that the integrand of (\ref{Klim}) becomes singular at $Y_{U,\alpha}=\{y_1, y_2, \dots, y_N\}$. The dashed integral represents the Cauchy principle value integral. 

Here, since $\alpha \in \mathcal{R}$, the set $Y_{U,\alpha}$ is non-empty. 
The terms in the summation in (\ref{Kri}) cannot cancel out because all the $M_{\alpha}^{-1}(y_j)$ are one-signed from the second Class (ii) condition. Also $U'(y_j)\neq 0$ for all $j$, as otherwise $M_{\alpha}^{-1}(y)$ cannot be continuous. Moreover, $\psi_0$ does not vanish in $\Omega$ because it is the least oscillatory eigenmode, as remarked just below \textcolor{black}{equation} (\ref{Rquo}). From (\ref{Kri}) $K_i$ is non-zero, and therefore we can conclude $\left. \frac{dc_i}{d\gamma}\right |_0\neq 0$ using (\ref{dcdg}). 

\textcolor{black}{We still need to show the latter half of Theorem \ref{unstable} for Class (i) basic flows. It is sufficient to consider the case $M_{\alpha}^{-1}(y)>0$ for all $y$, as otherwise the flow must be stable from Theorem 1. 
For any unstable mode $\psi$ we can deduce (\ref{aint}), which yields
$R_{\alpha}(\psi)<0$. Thus if we suppose there is no $g\in H^1_0$ such that $R_{\alpha}(g)<0$, then there should be no unstable modes.}


\subsection{Simple integral-based stability conditions}
If a function $g\in H^1_0$ that makes the Rayleigh quotient negative is found, the existence of instability is guaranteed by Theorem \ref{unstable}. 
The remaining question for deriving simple necessary conditions for stability is how to choose $g$. A good test function is $g=\varphi_0$, the function introduced in (\ref{kappa0}). 
Using (\ref{Rquo}), the condition $R_{\alpha}(\varphi_0)<0$ becomes
\begin{eqnarray}
1<\frac{2}{L}\int^{\frac{L}{2}}_{-\frac{L}{2}}M_{\alpha}^{-1} \cos^2 \left (\frac{\pi y}{L} \right) dy \, .
\label{DNC}
\end{eqnarray}
Note that this result depends on the boundary conditions. For example, if periodic boundary conditions are used, the eigenvalue problem (\ref{phiEQ}) produces (\ref{kappa0b}), and thus the condition (\ref{DNC}) must be replaced by
\begin{eqnarray}
1<\frac{1}{L}\int^{\frac{L}{2}}_{-\frac{L}{2}}M_{\alpha}^{-1}dy \, .
\label{perM}
\end{eqnarray}

\subsection{A simple stability condition using a hurdle}

The above conditions require us to examine the profile of $M^{-1}_{\alpha}(y)$ across the entire domain, $\Omega$. 
\textcolor{black}{However,} it turns out we can show the existence of instability by just checking a local part of the basic flow. Consider a comparison problem in a subinterval $[y_1,y_2] \subset \Omega$ with some $\widetilde{M}^{-1}(y)$ profile. 
\begin{eqnarray}
\widetilde{\psi}''-(\widetilde{k}^2+L_d^{-2})\widetilde{\psi}+\kappa_{0}^2\widetilde{M}^{-1}\widetilde{\psi}=0, \qquad y \in (y_1,y_2) \, .
\label{comparison}
\end{eqnarray}
We impose the Dirichlet condition, $\widetilde{\psi}(y_1)=\widetilde{\psi}(y_2)=0$. 
If this problem has a neutral mode, $\widetilde{\psi}_0(y)$ say, and $\widetilde{M}^{-1}<  M_{\alpha}^{-1}$ on $[y_1,y_2]$, the original problem (\ref{EQ2}) must be unstable. The reason is as follows. We set the test function as
\begin{equation}
g(y) = 
  \begin{cases}
    \widetilde{\psi}_0(y) \, , &  \text{if $y\in [y_1,y_2]$} \\ 
    0 \, ,                     & \text{otherwise}
  \end{cases} \label{testgg}
\end{equation}
using the neutral mode. Then the Rayleigh quotient can be computed as
\begin{eqnarray}
R(g)&=& \frac{\int_{y_1}^{y_2} |\widetilde{\psi}_0'|^2+(L_d^{-2}-\kappa_{0}^2M_{\alpha}^{-1})|\widetilde{\psi}_0|^2dy}{\int_{y_1}^{y_2} |\widetilde{\psi}_0|^2dy} \nonumber \\
&<& \frac{\int_{y_1}^{y_2} |\widetilde{\psi}_0'|^2+(L_d^{-2}-\kappa_{0}^2\widetilde{M}^{-1})|\widetilde{\psi}_0|^2dy}{\int_{y_1}^{y_2} |\widetilde{\psi}_0|^2dy}=-\widetilde{k}^2\leq 0 \, ,
\end{eqnarray}
where $\widetilde{k}$ is the wavenumber of the neutral mode $\widetilde{\psi}_0(y)$. 
A particularly convenient comparison problem is when $\widetilde{M}^{-1}$ is a constant, since the problem can be solved analytically using trigonometric functions.
It is easy to check that a neutral mode with a wavenumber $\widetilde{k}$ can be found when 
\begin{eqnarray}
\widetilde{M}^{-1}=\frac{\frac{\pi^2}{(y_2-y_1)^2}+(\widetilde{k}^2+L_d^{-2})}{\kappa_{0}^2} \, .
\end{eqnarray}
We want to make $\widetilde{M}^{-1}$ as small as possible to produce instability criteria as sharp as possible, so we let $\widetilde{k}\rightarrow 0$. Then the condition $\widetilde{M}^{-1}<  M_{\alpha}^{-1}$ on $[y_1,y_2]$ yields Theorem 2. 
\textcolor{black}{In brief, the hurdle theorem corresponds to selecting the test function $g(y)$ in (\ref{testgg}) with $\widetilde{\psi}_0(y)=\sin (\pi \frac{y-y_1}{y_2-y_1})$. It is possible to design test functions that provide greater sensitivity to the conditions, but in this article we have chosen to prioritise simplicity of the criteria.}



\section{Conclusion \textcolor{black}{and discussion}}
\label{sec:conclusion}
Rayleigh's inflection-point theorem of 1880 ushered in a century-long series of sufficient conditions for the stability of inviscid, parallel shear flows. Yet, well into the $21^{\rm st}$ century, 
 \textcolor{black}{sufficient conditions for instability (or equivalently, necessary conditions for stability)} 
have remained elusive for applied researchers. We have developed a simple methodology that allows for the detection of inviscid instability, inspired by empirical observations of the stability of Jupiter's alternating jets. All that is needed to operate our method is a comparison of $h$ defined in (\ref{hhh}) with the profile of the reciprocal Rossby Mach number, $M_{\alpha}^{-1}(y)$, which can be easily calculated from the basic flow (see section 3.3). The true stability boundary in the parameter space should be found between the unstable regions detected by Theorem 2 and the stable regions deduced by the KA criteria. 

Our method relies solely on the fundamental tools of Sturm-Liouville theory, but exhibits a high level of applicability beyond Rayleigh's equation framework. Similar hurdle stability conditions guaranteeing instability are likely to be found for various inviscid shear-flow stability problems in science and engineering, including hypersonic, stratified, magneto-hydrodynamic, and/or non-Newtonian flows. To the best of our knowledge, the possibility of such a general and practical instability theory has not been previously pointed out. 
\textcolor{black}{The hurdle theory could even yield new insights into classical Rayleigh equation problems as illustrated in Appendix C, where an internally heated vertical channel is analysed. 
}

As noted in section 1, in order for \textcolor{black}{the sufficient condition of instability (Theorem 2) to apply, the basic flow must belong to a class with certain favourable properties.} We assume there is at least one \textcolor{black}{PV extremum} in the domain and identify two basic-flow classes of interest, for which $M^{-1}_{\alpha}(y)$ is continuous in some reference zonal wind shift, $\alpha$. From the perspective of the general inviscid stability problem, one of the novelties of our work is the extension of the applicability of stability conditions through the introduction of new classifications. As illustrated by Theorem 5, our \textcolor{black}{sufficient condition of instability} apply to Class (ii), \textcolor{black}{which covers almost all monotonic flows and a wide range of non-monotonic flows.} If the basic flow belongs to this class, it can be asserted that there exists the \textcolor{black}{least oscillatory} regular neutral mode (section 6.3) and that instability occurs when the parameters are slightly altered (section 6.4). Theorem 5 requires a test function, but the stability condition can be simplified to \textcolor{black}{the hurdle form} using a kind of comparison principle (section 6.6). \textcolor{black}{A convenient method to determine whether the basic flow belongs to Class (ii) is as follows. (1) Plot $U(y)$ and mark all the points at the latitudes where $Q'=0$, $y=y_j$ say. Then for each point, draw a line at constant $U=U(y_j)$. If all of these lines intersect the $U(y)$ curve at points which are not marked, then the profile is not in Class (ii). (2) If the profile passes the first test, then check whether there is an odd number of zeros of $Q'$ in between any successive pair of zeros of $U-\alpha$. If so, this is also not in Class (ii). If the profile passes both tests, the profile is in Class (ii).}

In general, Class (ii) admits more than one reference-frame shift, $\alpha$ (see section 4.2). However, for a subclass of it, referred to as Class (i), $\alpha$ is uniquely determined (Theorem 3). For the latter class, $M^{-1}(y)$, which can now be safely unsubscripted, does not change sign. 
\textcolor{black}{Class (i) is approximately} identified with Jupiter and Saturn. \textcolor{black}{An important feature of Class (i) is that on the neutral curve the \textcolor{black}{PV extrema} must coincide with the critical \textcolor{black}{latitudes}, where the zonal wind speed matches with the phase speed of the neutral mode $\alpha$. 
Around the neutral parameters, vortices are produced near the critical latitudes, although mathematically we can show that all neutral modes have no singularity there.
Furthermore, for Class (i), Theorem 2 gives the necessary and sufficient condition for the stability when the limit $L_d\rightarrow 0$ is taken.}

\textcolor{black}{
The stability results for the linearised $1\frac{3}{4}$ layer model are summarised as follows:
\begin{itemize}
\item If $Q'$ has no zeros, the flow is stable (Charney-Stern).
\item Even if $Q'$ has zeros, if the flow is Class (i), it is stable if either $M^{-1}\leq 0$ everywhere (KA-I) or $M^{-1}\leq 1$ everywhere (KA-II).
\item If the flow is Class (ii), it is unstable if there are $\alpha, h$ such that $M_{\alpha}^{-1}> h$ in the interval where the hurdle $h$ is defined (Hurdle theorem).
\end{itemize}
\vspace{2mm}
If the flow is not Class (ii), then numerical methods are currently the primary recourse to determining stability.
}

Our results also offer new insights into the theoretical understanding of pattern formation in planetary atmospheres. The KA stability theorems were re-expressed by \cite{Dowling2014} in terms the Rossby Mach number, revealing a key attribute that supersonic critical latitudes are stable. However, this left the natural question of how subsonic a critical latitude must be before becoming unstable, which is answered precisely by Theorem 2, and illustrated with detailed model analyses in sections 4 and 5. 
In the case where $M^{-1}$ is a constant, the sonic condition, $M^{-1}=1$, must give the stability boundary, according to Theorems 1 and 2. 
Therefore, the conjecture of \cite{Stamp1993} is mathematically confirmed to be correct. 
In section 4 we extended the \cite{Stamp1993} model to study the case where $M^{-1}$ is not constant.  The $\rm sech^2$ bump profile considered for $M^{-1}(y)$ (see (\ref{Mprofile})) is interesting from the perspective of planetary physics because the behaviour of neutral curves can be analysed analytically to some extent. There exist qualitatively distinct neutral states, oscillatory and localised modes. In the context of Jupiter's atmosphere, only the latter appears, and the KA-II condition becomes almost sharp (figure 13). This implies that in atmospheres sustained over long periods, the $M^{-1}$ profile cannot become too subsonic, aligning with observational facts. However, a closer look in figure 1 reveals that the $Q_y$ curve sits above $U$ in the latitude range of -30 to -20 degrees. Unsteady dynamics may occur at neutral or subsonic critical latitudes there \citep[see][]{Read2006,Read2009}. 


From the $M^{-1}$ profile under the assumption of neutrality, in the context of $1$--$\nicefrac{3}{4}$ layer dynamics, we can calculate $U_{\rm deep}(y)$ from equation (\ref{Q'}). It is noteworthy that the estimate of $U_{\rm deep}(y)$ has been historically important for probing the deep jets on Jupiter \citep{Dowling1995_estimate}, though additional considerations for Jupiter in combination with Juno gravity inversions of interior circulations, and for Saturn in combination with ring-wave seismology (i.e., kronoseismology), are warranted in the future. 

The quasi-correlations of the zonal flow $U$ and the PV gradient $Q'$ seen in figure 1 suggests that the Jupiter's alternate jets are formed to be nearly linearly neutral with respect to inviscid Rossby wave instability. To understand why such a mean flow is achieved, it is necessary to consider the nonlinear evolution of the disturbances and their mutual interaction with the mean flow. It is numerically shown in \cite{Dowling1993} that when Jupiter's observed basic flow is made the initial condition for a local \textcolor{black}{unforced} shallow water model, it will rapidly evolve to become marginally stable. In addition, slowly moving, planetary-scale thermal features have been regularly observed on Jupiter \citep[][and references therein]{Fisher2016}. These lines of evidence suggest that a gas giant's alternating jets tend to evolve until the longest Rossby waves are coherent across them, becoming stationary with respect to deep-seated, large-scale pressure anomalies; this has been called a ``princess and the pea'' phenomenon \citep{Stamp1993}.

\textcolor{black}{The spontaneous generation mechanism of zonal jets has long remained a subject of debate among experts, as documented in \cite{Read2024}. The central inquiry lies in understanding how the energy provided by the heat from the sun or the interior of gas giants is converted into the kinetic energy of zonal and equatorial jets. While some numerical successes have emerged over the decade (see \cite{SchneiderLiu2009}, \cite{Lian2010}), a comprehensive and rational explanation is still lacking. Decomposing the flow into non-axisymmetric and axisymmetric components marks the first step in observing how Rossby waves transfer angular momentum within zonal mean flows, facilitated by Reynolds stress \citep{AndrewsMcIntyre1976, AndrewsMcIntyre1978}. The observed strong correlation between zonal flow and Reynolds stress suggests that this is indeed an indispensable mechanism (e.g. \cite{IngersollBeebe1981}). 
%
\textcolor{black}{Coupling the mean flow with linear perturbation equations through Reynolds flux terms is often called the mean-field approximation or quasi-linear theory in the modelling community (e.g. \cite{OGSc07}).}
It is noteworthy that studies on near-wall turbulence have demonstrated how the interaction between mean vortex fields and inviscid neutral waves rationally elucidates the generation of streaks \citep{Wang2007,HallSherwin2010}. Recent investigations have further pinpointed the presence of neutral waves within large-scale laminar-turbulent patterns in rotating flows \citep{Wang2022}.
Thus, we expect that the formation and evolution of Jupiter's alternating jets might be similarly explored by coupling the mean flow equations for the planet's atmosphere and interior with the inviscid stability problem of the mean flow, along the lines of similar theories developed in near-wall turbulence. 
}

\bigskip
Acknowledgement. \textcolor{black}{We thank the anonymous referees for their constructive comments that helped to improve the article.} The data for figures~\ref{fig:Read2006_profiles} and \ref{fig:Read2006_profiles2}  were kindly provided by P.L. Read.\\

Declaration of interests. The authors report no conflict of interest.\\

\appendix

\section{Sufficient conditions of stability}

Consider the integral of  $\psi^*\times$(\ref{EQ}) over $\Omega=(-L/2,L/2)$. Integration by parts yields
\begin{eqnarray}
\int_{\Omega} \frac{Q'(U-c^*)}{|U-c|^2} |\psi|^2 dy=\int_{\Omega} \{|\psi'|^2+(k^2+L_d^{-2})|\psi|^2\} dy \, ,
\label{intstab}
\end{eqnarray}
whose real and imaginary parts are
\begin{eqnarray*}
\int_{\Omega}  \frac{Q'(U-c_r)}{|U-c|^2} |\psi|^2 dy=\int_{\Omega}  \{|\psi'|^2+(k^2+L_d^{-2})|\psi|^2\} dy,~~~
c_i\int_{\Omega}  \frac{Q'}{|U-c|^2} |\psi|^2 dy=0 \, ,
\end{eqnarray*}
respectively.
When we assume $c_i \ne 0$, the above two equations can be combined into
\begin{eqnarray}
\int_{\Omega} \frac{Q'(U-w)}{|U-c|^2} |\psi|^2 dy=\int_{\Omega}  \{|\psi'|^2+(k^2+L_d^{-2})|\psi|^2\} dy \, ,
\label{wint}
\end{eqnarray}
where $w\in \RealN$ is arbitrary. 
If there exists $w$ such that 
$Q'/(U-w) \leq 0$
for all $y$, (\ref{wint}) cannot be satisfied, meaning that the flow is stable. In other words, the stability is guaranteed if there exists $w \in \RealN$ such that $M_{w}^{-1}\leq 0$ for all $y\in \Omega$ (i.e. KA-I). The above derivation is essentially the standard argument for shear flows due to Rayleigh and Fj{\o}rtoft.

To find KA-II, we rewrite the left hand side of (\ref{wint}) by introducing an $\alpha\in \RealN$ and setting $w=2c_r-\alpha$:
\begin{eqnarray}
\int_{\Omega} \frac{(U-w)(U-\alpha)}{(U-c_r)^2+c_i^2}\frac{Q'}{U-\alpha} |\psi|^2 dy=\int_{\Omega} \frac{
(U-c_r)^2-(c_r -\alpha)^2
}{(U-c_r)^2+c_i^2}\frac{Q'}{U-\alpha} |\psi|^2 dy \, . ~~~~
\end{eqnarray}
Using Poincare's inequality $\frac{L^2}{\pi^2}\int_{\Omega}|\phi'|^2 dy\geq \int_{\Omega}|\phi|^2 dy$, from (\ref{wint}) we can deduce
\begin{eqnarray}
\kappa_0^2\int_{\Omega} ZM_{\alpha}^{-1} |\psi|^2 dy=\int_{\Omega}  \{|\psi'|^2+(k^2+L_d^{-2})|\psi|^2\} dy\geq (\kappa_0^2+k^2)\int_{\Omega} |\psi|^2 dy,~~~~\label{aint}
\end{eqnarray}
where $Z(y)=\frac{
(U-c_r)^2-(c_r -\alpha)^2
}{(U-c_r)^2+c_i^2}<1$ for all $y$.
If $M_{\alpha}^{-1}\in [0,1]$ for all $y$ the inequality (\ref{aint}) cannot be satisfied, so the flow must be stable. 
~\\

\section{Analysis of (\ref{nEQ}) for $L\gg 1$}
Writing 
\begin{eqnarray}
\nu=\frac{\sqrt{1+4\kappa_{0}^2(b-a)}-1}{2}, \quad \mu=\sqrt{k^2+L_d^{-2}-\kappa_{0}^2a} \, ,\label{numu}
\end{eqnarray}
the general solution of (\ref{nEQ}) can be written as
\begin{eqnarray}
\psi(y)=C_1P^{\mu}_{\nu}(\text{tanh}y)+C_2Q^{\mu}_{\nu}(\text{tanh}y)
\end{eqnarray}
using arbitrary constants $C_1$ and $C_2$. 
Here
$P^{\mu}_{\nu}$ and $Q^{\mu}_{\nu}$ are the associated Legendre functions of the first and second kind, respectively.
The constant $\mu$ determines the behaviour of the solution at large $|y|$. If $\mu$ is real, the behaviour is exponential, while if $\mu$ is imaginary, the behaviour is oscillatory (see \cite{Bielski2013}, for example). Now we assume that $L$ is large. Since $\kappa_0^2\approx L_d^{-2}$ under this assumption, the constant $\mu$ can become imaginary only when $a>1$. This is the case where the oscillatory modes may appear.

If $\mu$ and $\nu$ are integers satisfying $0\leq \mu \leq \nu$, the neutral mode can be obtained by using the Legendre polynomials. The localised mode $\psi=\text{sech}y$ found in section 5.2 is the special case $\nu=\mu=1$. The parabolic part of the neutral curve (\ref{Tneutral}) can be found by generalising this mode, because from Sturm's oscillation theory, the mode without \textcolor{black}{zeros}, i.e. $\psi_0$, is usually the most dangerous. This mode can be written explicitly as $\psi_0(y)=(\text{sech}y)^{\nu}$, because it satisfies (\ref{nEQ}) when $\nu=\mu$. For the solution to decay as $|y|\rightarrow \infty$, it is necessary for $\nu$ to be greater than 0, a condition that is indeed satisfied in the forth quadrant of figure 6a. 
The condition $\nu=\mu$ can be written as
\begin{eqnarray}
k^2=\frac{1}{2}\left(1+2L_d^{-2}(b-1)-\sqrt{1+4L_d^{-2}(b-a)} \right )
\label{KLBA}
\end{eqnarray}
using (\ref{numu}). Thus the neutral mode $\psi_0$ exists when the right hand side of (\ref{KLBA}) is non-negative. After some algebra, this condition can be simplified to
\begin{eqnarray}
a\geq 1-L^{-2}_d(b-1)^2 \, .
\end{eqnarray}

\section{The Rayleigh equation case: internally heated flow}
\begin{figure}
\begin{center}
  \includegraphics[width=0.9\textwidth]{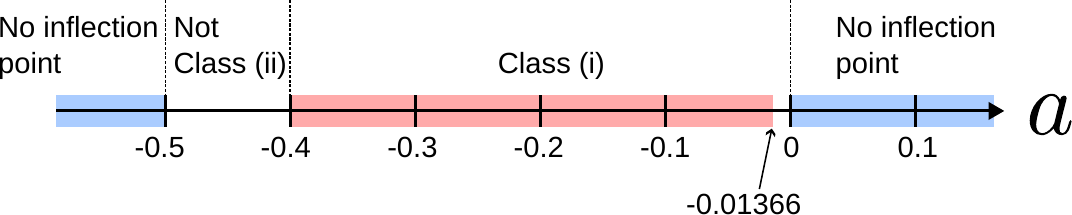}
\end{center}
\caption{The stability of the Rayleigh equation with the model flow profile (\ref{IntH}). The blue regions are KA stable because there are no inflection points in $U(y)$. The red region is unstable according to Theorem \ref{ThmM}. \textcolor{black}{Numerical computations detect instability for $a\in(-0.5,0)$.}
}
\label{fig4}
\end{figure}

Here we consider the Rayleigh equation case, i.e., $Q'=-U''$ and $L_d^{-2}=0$ in (\ref{EQ}). 
Our stability condition (Theorem 2) has broader applicability compared to existing ones, however, there may be cases where existing conditions are more suitable when they can be applied. To demonstrate this, here we purposefully choose a basic flow for which the condition by \cite{Tollmien1935} is applicable.

The basic flow we choose is
\begin{eqnarray}
U(y) = \frac{y^4-6y^2+5}{12}+a(1-y^2) \, ,
\label{IntH}
\end{eqnarray}
which appears when internally heated fluid flows through a vertically oriented channel; see \cite{Nagata_Generalis2002}. The walls are set at $y=-1$ and 1, hence $L=2$. The parameter $a$ is the ratio of the Reynolds number to the Grashof number. Based on the Orr-Sommerfeld numerical calculations, \cite{Uhlmann_Nagata2006} argued that for $O(1)$ wavenumbers, instability appears when $a \in (-5/12,0)\approx (-0.4166,0)$, because of the existence of inflection points and reverse flow. Note that unstable modes also exists at small wavenumber parameter regions due to a viscous instability mechanism (i.e., Tollmien–Schlichting waves), but that is not the current focus. 

The basic flow (\ref{IntH}) has two inflection points in $\Omega=(-1,1)$ when $a \in (-1/2,0)$, $y_c=\pm \sqrt{2a+1}$. For other values of $a$, there are no inflection points and the flow must be stable according to KA-I; see figure \ref{fig4}. Using $\alpha=U(y_c)=-a(5a+2)/3$, 
the reciprocal Rossby Mach number can be obtained as
\begin{eqnarray}
M^{-1}_{\alpha}(y) =\frac{L^2}{\pi^2}\frac{-U''}{U-\alpha}= \frac{1}{\pi^2}\frac{48}{10a+5-y^2} \, .
\end{eqnarray}
This is continuous in $\Omega=(-1,1)$ when $a\geq -2/5$, so we conclude that the flow is Class (i) if $a \in [-2/5,0)$. In the remaining parameter range $a \in (-1/2,-2/5)$, there are no $\alpha \in \mathcal{R}$ for which $M^{-1}_{\alpha}(y)$ is continuous, and hence the basic flow is not Class (ii) (and thus not Class (i)).

For $a \in  [-2/5,0)$ we can use Theorem \ref{ThmM} to show that the flow is unstable when $a\in [-\frac{2}{5}, \frac{24}{\pi^2}-\frac{1}{2})\approx [-0.4,  -0.01366)$. This result can be found by simply setting $y_2=1, y_1=-1$; in this case the height of the hurdle is 1 according to (\ref{hhh}). Note that the stability condition (\ref{C1KA}) is satisfied when $a$ is greater than $\frac{24}{\pi^2}-\frac{2}{5}\approx 0.08634$, but it is not useful because the flow is not Class (i) there, due to the absence of critical latitudes (inflection points). 

If $a\notin (-1/2,-1/3)$, the basic flow belongs to the class considered in \cite{Tollmien1935}, and therefore the KA-I condition provides a sharp stability boundary at $a=0$. Thus Tollmien's theory more accurately pinpoints the stability boundary than (\ref{C1KA}). However, in terms of applicability, the advantage is reversed, since $(-1/2,-2/5)\subset (-1/2,-1/3)$ (see figure \ref{fig4}). In fact, Tollmien's result can be obtained by taking $\psi=U(y)$ as a test function, which is only possible for very special class of basic flows. 
Note also that to identify instability in the current problem, focusing on the domain-spanning hurdle is sufficient. Drazin \& Howard (1966) have previously investigated the existence of neutral eigensolutions in this scenario. However, demonstrating the presence of instability requires proper treatment of multiple critical layers in the flow.

\begin{figure}
\begin{center}
 \includegraphics[width=0.5\textwidth]{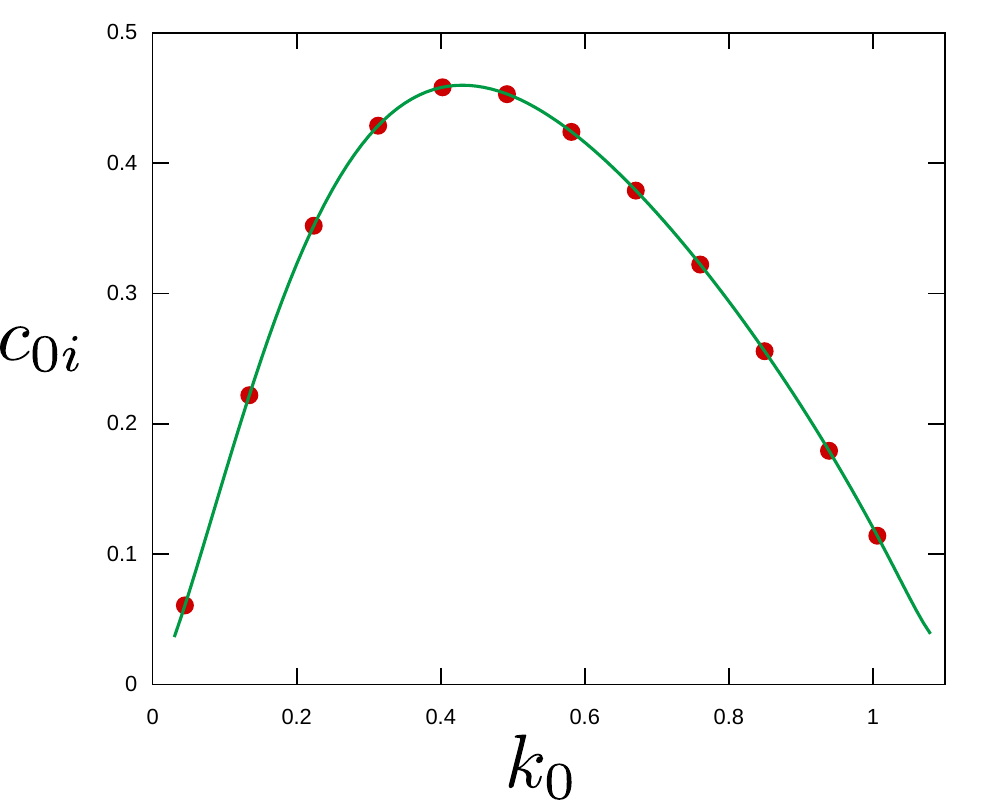}
\end{center}
\caption{The green curve is the solution of the asymptotic problem (\ref{asym05}). The red filled circles are the solution of the Rayleigh equation with the basic flow (\ref{IntH}). 
The growth rate $c_i$ computed for various $k$ for $a=-0.45$, and then the data are rescaled as $c_{0i}=c_i/\epsilon^4$, $k_0=k/\epsilon$, using $\epsilon=\sqrt{0.5-0.45}\approx 0.2236$.
}
\label{fig11}
\end{figure}

Numerical computations using the shooting or Chebyshev collocation method reveal that the flow is unstable when $a\in (-0.5,0)$. 
\textcolor{black}{Consequently, the inference made by \cite{Uhlmann_Nagata2006} was incorrect.}
The computation becomes challenging around $a=-0.5$, but we can employ asymptotic methods to address this.
Let us consider $\epsilon=\sqrt{a+\frac{1}{2}}$ as a small parameter. Then we write $y=\epsilon Y$, $k=\epsilon^{-1}k_0$ and
\begin{eqnarray}
c=-\frac{1}{12}+\epsilon ^{2} +\epsilon^{4}c_0+\cdots
\end{eqnarray}
in the Rayleigh equation. 
Noting $U-c =\epsilon ^{4}(\frac{Y^4}{12}-Y^2-c_0)+\cdots$ and $U''=\epsilon ^{2} (Y^2- 2)+\cdots$, 
the leading order equation can be found as
\begin{equation}
\left (\frac{Y^4}{12}-Y^2-c_0 \right)(\psi_{YY}-k_0^2\psi) -(Y^2- 2)\psi=0.\label{asym05}
\end{equation}
Here we must impose $\psi\rightarrow 0$ as $|Y|\rightarrow \infty$. This eigenvalue problem can be solved by the shooting method to find the eigenvalue $c_0=c_{0r}+ic_{0i}$ for fixed $k_0$. The asymptotic growth rate $c_{0i}$, shown in figure \ref{fig11}, captures the behaviour of the full solution very well even when $\epsilon$ is moderately small. 
\textcolor{black}{Since this instability appears around singular neutral modes, it is outside the application of the sufficient conditions of stability available so far.}


Whether simple stability criteria can be obtained for basic flows that do not belong to Class (ii) is still an open question. A central issue here is determining under what conditions singular neutral modes must occur.
Due to the presence of jumps at singular points, $\lambda=-k^2$ is no longer obtained as an eigenvalue of a self-adjoint operator, and the optimisation of a quadratic form (\ref{Rquo}) cannot be used to demonstrate the presence of neutral modes. 


Another open problem is extension of the theory to basic flows that vary spatially in a more complicated manner than considered here. In other shear flow studies, despite the change of the basic flow in two directions being the prevalent configuration in almost all realistic flows, the Rayleigh-Fj{\o}rtoft condition, which only holds in idealised situations, is often employed to explain the origin of instability. A typical example is the Kelvin-Helmholtz instability observed in flows over a riblet \citep[e.g.][]{Garcia-Mayoral_Jimenez2011}. 
\cite{Uhlmann_Nagata2006} also studied inviscid instability in duct flows using second derivatives of the basic flow along the steepest direction of the flow field. Curiously, they found that the presence of  inflection points calculated in this manner is effective in detecting instability. This is further evidence that the method developed in this paper could be generalised. A difficulty with this extension is that the necessary conditions for the existence of a neutral mode are not known, although a sufficient condition has recently been found in the case of a single critical layer \citep{Deguchi2019}.

\bibliographystyle{jfm}  
\bibliography{main}  

\end{document}